\documentclass[letterpaper,journal]{IEEEtran}
\IEEEoverridecommandlockouts
\usepackage[T1]{fontenc}
\usepackage{lmodern}
\usepackage{cite}
\usepackage{amsmath,amssymb,amsfonts,amsthm,bm,mathrsfs,mathtools}
\usepackage{algorithmic}
\usepackage{algorithm}
\usepackage{amsbsy}
\usepackage{epsf}
\usepackage{colortbl}
\usepackage{color}
\definecolor{lightgrey}{rgb}{0.8,0.8,0.8}
\definecolor{lightblue}{rgb}{0.68, 0.85, 0.9}
\usepackage[export]{adjustbox}
\usepackage[shortlabels]{enumitem}
\usepackage{soul}
\usepackage{tabularx}
\usepackage[english]{babel}
\addto\captionsenglish{\renewcommand{\figurename}{Fig.}}
\usepackage{graphicx}
\usepackage{textcomp}
\usepackage{xcolor}
\def\BibTeX{{\rm B\kern-.05em{\sc i\kern-.025em b}\kern-.08em
    T\kern-.1667em\lower.7ex\hbox{E}\kern-.125emX}}

\newcounter{mbcomment}
\newcounter{mbcommentr}
\newcounter{mbcommento}
\newcounter{mbnote}

\usepackage{nicematrix}

\usepackage{comment}
\usepackage{glossaries}
\usepackage{quotes}

\newcommand{\smily}[1]{\overset{\lower0.25em\hbox{$\smash{\scriptscriptstyle\smile}$}}{#1}}
\newcommand{\frowny}[1]{\overset{\lower0.25em\hbox{$\smash{\scriptscriptstyle\frown}$}}{#1}}
\newcommand{\plusy}[1]{\overset{\lower0.25em\hbox{$\smash{\scriptscriptstyle +}$}}{#1}}
\newcommand{\circly}[1]{\overset{\lower0.25em\hbox{$\smash{\scriptscriptstyle \circ}$}}{#1}}
\newcommand{\overy}[2]{\overset{\lower0.25em\hbox{$\smash{\scriptscriptstyle #2}$}}{#1}}

\definecolor{titles}{HTML}{CC00FF}
\definecolor{subtitles}{HTML}{E97132}
\definecolor{math}{HTML}{0066FF}
\definecolor{DarkRosyBrown}{RGB}{176, 84, 48}
\definecolor{BGPink}{RGB}{244, 224, 244}
\definecolor{BGBrown}{RGB}{228, 204, 184}
\definecolor{BGGreen}{HTML}{CAEEF3}
\definecolor{BGBlue}{HTML}{66FFFF}
\definecolor{BrightGreen}{RGB}{44, 164, 48}

\newcounter{DraftBox}

\newcounter{QuestionBox}

\usepackage{tabularx,makecell,booktabs,multirow}

\usepackage{adjustbox} 
\usepackage{pgfplots}
\pgfplotsset{compat=1.17}



\usepackage{pifont}    

\newcommand{\xmark}{\ding{55}}%

\usepackage{amssymb}   
\usepackage{makecell}  
\newacronym{1g}{1G}{first generation}
\newacronym{2g}{2G}{second generation}
\newacronym{3g}{3G}{third generation}
\newacronym{3gpp}{3GPP}{3rd Generation Partnership Project}
\newacronym{4g}{4G}{fourth generation}
\newacronym{5g}{5G}{fifth generation}
\newacronym{5gb}{5G\&B}{\gls{5g} and beyond}
\newacronym{b5g}{B5G}{beyond fifth generation}
\newacronym{6g}{6G}{sixth generation}
\newacronym{lte}{LTE}{long-term evolution}
\newacronym{ltem}{LTE-M}{long-term evolution for mobile}
\newacronym{nr}{NR}{new radio}
\newacronym{gsm}{GSM}{Global System for Mobile Communications}
\newacronym{umts}{UMTS}{Universal Mobile Telecommunications System}
\newacronym{wifi}{Wi-Fi}{wireless fidelity}
\newacronym{r18}{3GPP Release 18}{3GPP Release 18}

\newacronym{adc}{ADC}{analog-to-digital converter}
\newacronym{dac}{DAC}{digital-to-analog converter}
\newacronym{ap}{AP}{access point}
\newacronym{bs}{BS}{base station}
\newacronym{ue}{UE}{user equipment}
\newacronym{ris}{RIS}{reconfigurable intelligent surface}
\newacronym{star}{STAR}{simultaneously transmitting and reflecting}
\newacronym{das}{DAS}{distributed antenna system}
\newacronym{mbs}{MBS}{macro base station}
\newacronym{nap}{NAP}{network access point}
\newacronym{rau}{RAU}{remote access unit}
\newacronym{uav}{UAV}{unmanned aerial vehicle}
\newacronym{gs}{GS}{ground station}

\newacronym{ula}{ULA}{uniform linear array}
\newacronym{upa}{UPA}{uniform planar array}
\newacronym{siso}{SISO}{single-input single-output}
\newacronym{miso}{MISO}{multiple-input single-output}
\newacronym{simo}{SIMO}{single-input multiple-output}
\newacronym{mimo}{MIMO}{multiple-input multiple-output}
\newacronym{mmimo}{mMIMO}{massive multiple-input multiple-output}
\newacronym{cfmm}{CFMM}{cell-free massive MIMO}

\newacronym{led}{LED}{light emitting diode}
\newacronym{ld}{LD}{laser diode}
\newacronym{pd}{PD}{photodetector}
\newacronym{fbg}{FBG}{fiber Bragg gratings}
\newacronym{smf}{SMF}{single-mode fiber}
\newacronym{mmf}{MMF}{multimode fiber}
\newacronym{dcf}{DCF}{dispersion-compensated fiber}
\newacronym{dff}{DFF}{dispersion-flattened fiber}
\newacronym{dsf}{DSF}{dispersion-shifted fiber}
\newacronym{soa}{SOA}{semiconductor optical amplifier}

\newacronym{ofdm}{OFDM}{orthogonal frequency division multiplexing}
\newacronym{cpofdm}{CP-OFDM}{cyclic prefix-based OFDM}
\newacronym{dft}{DFT}{discrete Fourier transform}
\newacronym{idft}{IDFT}{inverse discrete Fourier transform}
\newacronym{fft}{FFT}{fast Fourier transform}
\newacronym{ifft}{IFFT}{inverse fast Fourier transform}
\newacronym{sfft}{SFFT}{symplectic finite Fourier transform}
\newacronym{isfft}{ISFFT}{inverse symplectic finite Fourier transform}
\newacronym{wht}{WHT}{Walsh-Hadamard transform}
\newacronym{cdma}{CDMA}{code division multiple access}
\newacronym{noma}{NOMA}{non-orthogonal multiple access}
\newacronym{tdma}{TDMA}{time division multiple access}
\newacronym{wdma}{WDMA}{wavelength division multiple access}
\newacronym{rsma}{RSMA}{rate-splitting multiple access}
\newacronym{scs}{SCS}{subcarrier spacing}
\newacronym{cp}{CP}{cyclic prefix}
\newacronym{zp}{ZP}{zero padding}

\newacronym{qam}{QAM}{quadrature amplitude modulation}
\newacronym{bpsk}{BPSK}{binary phase shift keying}
\newacronym{bfsk}{BFSK}{binary frequency shift keying}
\newacronym{pam}{PAM}{pulse amplitude modulation}
\newacronym{pwm}{PWM}{pulse width modulation}
\newacronym{oqam}{OQAM}{offset quadrature amplitude modulation}
\newacronym{im}{IM}{index modulation}
\newacronym{scm}{SCM}{single carrier modulations}
\newacronym{fbmc}{FBMC}{filter bank multicarrier}
\newacronym{gfdm}{GFDM}{generalized frequency division multiplexing}
\newacronym{ufmc}{UFMC}{universal filtered multicarrier}
\newacronym{otfs}{OTFS}{orthogonal time frequency space}
\newacronym{otsm}{OTSM}{orthogonal time sequency multiplexing}
\newacronym{ocdm}{OCDM}{orthogonal chirp division multiplexing}

\newacronym{vlc}{VLC}{visible light communications}
\newacronym{owc}{OWC}{optical wireless communications}
\newacronym{fso}{FSO}{free space optical}
\newacronym{imdd}{IM/DD}{intensity modulation with direct detection}
\newacronym{acoofdm}{ACO-OFDM}{asymmetrically clipped optical OFDM}
\newacronym{adoofdm}{ADO-OFDM}{asymmetrically clipped DC biased optical OFDM}
\newacronym{coofdm}{CO-OFDM}{coherent optical OFDM}
\newacronym{cqoofdm}{CQO-OFDM}{complex QAM optical OFDM}
\newacronym{dcoofdm}{DCO-OFDM}{DC biased optical OFDM}
\newacronym{ddoofdm}{DDO-OFDM}{direct detection optical OFDM}
\newacronym{lacoofdm}{LACO-OFDM}{layered asymmetrically clipped optical OFDM}
\newacronym{oofdm}{O-OFDM}{optical OFDM}
\newacronym{scoofdm}{SCO-OFDM}{self-coherent optical OFDM}

\newacronym{dsp}{DSP}{digital signal processing}
\newacronym{csi}{CSI}{channel state information}
\newacronym{rss}{RSS}{received signal strength}
\newacronym{aoa}{AoA}{angle of arrival}
\newacronym{aod}{AoD}{angle of departure}
\newacronym{los}{LoS}{line-of-sight}
\newacronym{nlos}{NLOS}{non-line-of-sight}
\newacronym{cir}{CIR}{channel impulse response}
\newacronym{isi}{ISI}{intersymbol interference}
\newacronym{ici}{ICI}{interchannel interference}
\newacronym{mai}{MAI}{multiple access interference}
\newacronym{ini}{INI}{inter-numerology interference}
\newacronym{sni}{SNI}{same numerology interference}
\newacronym{awgn}{AWGN}{additive white Gaussian noise}
\newacronym{sic}{SIC}{successive interference cancellation}
\newacronym{bem}{BEM}{basis expansion model}
\newacronym{svd}{SVD}{singular value decomposition}
\newacronym{apf}{APF}{all-pass filter}
\newacronym{ptf}{PTF}{prototype filter}
\newacronym{fir}{FIR}{finite-duration impulse response}
\newacronym{iir}{IIR}{infinite impulse response}

\newacronym{iot}{IoT}{internet of things}
\newacronym{nbiot}{NB-IoT}{narrowband internet of things}
\newacronym{iiot}{IIoT}{industrial internet of things}
\newacronym{oiot}{OIoT}{optical internet of things}
\newacronym{ioat}{IoAT}{internet of autonomous things}
\newacronym{mtc}{MTC}{machine-type communications}
\newacronym{mmtc}{mMTC}{massive machine-type communications}
\newacronym{m2m}{M2M}{machine-to-machine}
\newacronym{d2d}{D2D}{device-to-device}
\newacronym{v2x}{V2X}{vehicle-to-everything}
\newacronym{v2v}{V2V}{vehicle-to-vehicle}
\newacronym{v2i}{V2I}{vehicle-to-infrastructure}
\newacronym{v2p}{V2P}{vehicle-to-pedestrian}
\newacronym{hetnet}{HetNet}{heterogeneous network}
\newacronym{udn}{UDN}{ultra dense network}
\newacronym{sdn}{SDN}{software-defined network}
\newacronym{nfv}{NFV}{network function virtualization}
\newacronym{ran}{RAN}{radio access network}
\newacronym{wan}{WAN}{wide area network}
\newacronym{man}{MAN}{metropolitan area network}
\newacronym{lan}{LAN}{local area network}
\newacronym{wlan}{WLAN}{wireless local area network}
\newacronym{wsn}{WSN}{wireless sensor network}
\newacronym{manet}{MANET}{mobile ad hoc network}
\newacronym{tn}{TN}{terrestrial network}
\newacronym{ntn}{NTN}{non-terrestrial network}
\newacronym{is}{IS}{integrated satellite}
\newacronym{sagsin}{SAGSIN}{space-air-ground-sea integrated network}
\newacronym{atg}{ATG}{air-to-ground}
\newacronym{u2n}{U2N}{UAV-to-network}

\newacronym{embb}{eMBB}{enhanced mobile broadband}
\newacronym{urllc}{URLLC}{ultra-reliable low-latency communications}
\newacronym{mbb}{MBB}{mobile broadband}
\newacronym{its}{ITS}{intelligent transportation systems}
\newacronym{cras}{CRAS}{connected robotics and autonomous systems}
\newacronym{swipt}{SWIPT}{simultaneous wireless information and power transfer}
\newacronym{jcas}{JCAS}{joint communication and sensing}

\newacronym{cc}{CC}{cloud computing}
\newacronym{ec}{EC}{edge computing}
\newacronym{fc}{FC}{fog computing}
\newacronym{mec}{MEC}{multi-access edge computing}

\newacronym{ai}{AI}{artificial intelligence}
\newacronym{ml}{ML}{machine learning}
\newacronym{dlearning}{DL}{deep learning}
\newacronym{rl}{RL}{reinforcement learning}
\newacronym{drl}{DRL}{deep reinforcement learning}
\newacronym{ann}{ANN}{artificial neural network}
\newacronym{cnn}{CNN}{convolutional neural network}
\newacronym{dnn}{DNN}{deep neural network}
\newacronym{mlp}{MLP}{multilayer perceptron}
\newacronym{hsudnn}{HSUDNN}{hybrid semi-unfolding deep neural network}
\newacronym{baim}{BAIM}{big artificial intelligence model}

\newacronym{mdp}{MDP}{Markov decision process}
\newacronym{dqn}{DQN}{deep Q-network}
\newacronym{ddqn}{DDQN}{double deep Q-network}
\newacronym{sac}{SAC}{soft actor-critic}
\newacronym{ddpg}{DDPG}{deep deterministic policy gradient}
\newacronym{td3}{TD3}{twin delayed deep deterministic policy gradient}
\newacronym{sd3}{SD3}{smoothed softmax dual deep deterministic policy gradient}
\newacronym{ppo}{PPO}{proximal policy optimization}
\newacronym{dsl}{DSL}{differentiable safety layer}

\newacronym{ao}{AO}{alternating optimization}
\newacronym{bcd}{BCD}{block coordinate descent}
\newacronym{sca}{SCA}{successive convex approximation}
\newacronym{saa}{SAA}{sample average approximation}
\newacronym{sdr}{SDR}{semidefinite relaxation}
\newacronym{sdp}{SDP}{semidefinite programming}
\newacronym{qcqp}{QCQP}{quadratically constrained quadratic program}
\newacronym{sp}{SP}{S-procedure}
\newacronym{rcg}{RCG}{Riemannian conjugate gradient}
\newacronym{mmse}{MMSE}{minimum mean-square error}
\newacronym{wmmse}{WMMSE}{weighted minimum mean-square error}
\newacronym{ls}{LS}{least squares}
\newacronym{cnum}{CN}{concatenation number}
\newacronym{cn}{CN}{$\mathcal{CN}$—complex normal distribution}
\newacronym{iid}{i.i.d.}{independent and identically distributed}
\newacronym{pdf}{PDF}{probability density function}
\newacronym{cdf}{CDF}{cumulative distribution function}
\newacronym{mgf}{MGF}{moment generating function}
\newacronym{areg}{AR}{autoregressive}
\newacronym{arma}{ARMA}{autoregressive moving average}

\newacronym{qos}{QoS}{quality of service}
\newacronym{qoe}{QoE}{quality of experience}
\newacronym{sinr}{SINR}{signal-to-interference-plus-noise ratio}
\newacronym{snr}{SNR}{signal-to-noise ratio}
\newacronym{ber}{BER}{bit error rate}
\newacronym{aber}{ABER}{average bit error rate}
\newacronym{bep}{BEP}{bit error probability}
\newacronym{se}{SE}{spectral efficiency}
\newacronym{ee}{EE}{energy efficiency}
\newacronym{re}{RE}{resource efficiency}
\newacronym{pe}{PE}{power efficiency}
\newacronym{aoi}{AoI}{age of information}
\newacronym{papr}{PAPR}{peak-to-average power ratio}
\newacronym{mse}{MSE}{mean-square error}
\newacronym{ms}{MS}{mean square}
\newacronym{acp}{ACP}{average coverage probability}
\newacronym{sop}{SOP}{secrecy outage probability}
\newacronym{st}{ST}{secrecy throughput}
\newacronym{kpi}{KPI}{key performance indicator}

\newacronym{vr}{VR}{virtual reality}
\newacronym{ar}{AR}{augmented reality}
\newacronym{mr}{MR}{mixed reality}
\newacronym{xr}{XR}{extended reality}
\newacronym{gps}{GPS}{global positioning system}
\newacronym{gis}{GIS}{geographic information system}
\newacronym{slam}{SLAM}{simultaneous localization and mapping}
\newacronym{rfid}{RFID}{radio frequency identification}
\newacronym{nfc}{NFC}{near-field communication}
\newacronym{usb}{USB}{universal serial bus}

\newacronym{pls}{PLS}{physical layer security}
\newacronym{aes}{AES}{Advanced Encryption Standard}
\newacronym{des}{DES}{Data Encryption Standard}
\newacronym{rsa}{RSA}{Rivest-Shamir-Adleman}
\newacronym{dos}{DoS}{denial of service}
\newacronym{ddos}{DDoS}{distributed denial of service}
\newacronym{mitm}{MitM}{man-in-the-middle}

\newacronym{gym}{Gymnasium}{Gymnasium RL interface}
\newacronym{sbthree}{SB3}{Stable-Baselines3}
\newacronym{acm}{ACM}{Association for Computing Machinery}
\newacronym{iec}{IEC}{International Electrotechnical Commission}
\newacronym{nist}{NIST}{National Institute of Standards and Technology}
\newacronym{eh}{EH}{energy harvesting}
\newacronym{em}{EM}{electromagnetic}
\usepackage{hyperref}
\hypersetup{
    colorlinks=true,
    linkcolor=black,
    filecolor=magenta,      
    urlcolor=blue,
    citecolor=black,
    pdftitle={Overleaf Example},
    pdfpagemode=FullScreen,
    }
\definecolor{light_grey}{rgb}{0.8,0.8,0.8}
\definecolor{light_blue}{rgb}{0.68, 0.85, 0.9}
\usepackage{tcolorbox}
\definecolor{tableheads}{RGB}{244, 224, 244}
\usepackage{multirow}
\usepackage{float}
\usepackage{scalerel}
\title{Throughput Optimization in UAV-Mounted RIS under Jittering and Imperfect CSI via DRL}
\author{\IEEEauthorblockN{}
\IEEEauthorblockA{
Department of Electrical Engineering, King Fahd University of Petroleum and Minerals}}
\author{Anas K. Saeed, Mahmoud M. Salim, Ali Arshad Nasir, \IEEEmembership{Senior Member, IEEE}, \\
and Ali H. Muqaibel,
\IEEEmembership{Senior Member, IEEE}        
\thanks{All authors are with the Center for Communication Systems and Sensing, King Fahd University of Petroleum and Minerals, Dhahran 31261, Saudi Arabia. Also, Anas K. Saeed, Ali Arshad Nasir, and Ali H. Muqaibel are with the Electrical Engineering Department at King Fahd University of Petroleum and Minerals, Dhahran 31261, Saudi Arabia. Ali H. Muqaibel is the corresponding author (email: muqaibel@kfupm.edu.sa).}}

\begin{document}

\maketitle
\begin{abstract}
\Glspl{ris} mounted on \glspl{uav} can reshape wireless propagation on-demand. However, their performance is sensitive to \gls{uav} jitter and cascaded channel uncertainty. This paper investigates a downlink multiple-input single-output UAV-mounted \gls{ris} system in which a ground multiple-antenna \gls{bs} serves multiple single-antenna users under practical impairments. Our goal is to maximize the expected throughput under stochastic three-dimensional \gls{uav} jitter and imperfect cascaded \gls{csi} based only on the available channel estimates. This leads to a stochastic nonconvex optimization problem subject to a \gls{bs} transmit power constraint and strict unit-modulus constraints on all \gls{ris} elements. To address this problem, we design a model-free \gls{drl} framework with a contextual bandit formulation. A differentiable feasibility layer is utilized to map continuous actions to feasible solutions, while the reward is a Monte Carlo estimate of the expected throughput. We instantiate this framework with constrained variants of \gls{ddpg} and \gls{td3} that do not use target networks. Simulations show that the proposed algorithms yield higher throughput than conventional alternating optimization-based weighted minimum mean-square error (AO-WMMSE) baselines under severe jitter and low \gls{csi} quality. Across different scenarios, the proposed methods achieve performance that is either comparable to or slightly below the AO-WMMSE benchmark, based on sample average approximation (SAA) with a relative gap ranging from $0$-$12\%$. Moreover, the proposed \gls{drl} controllers achieve online inference times of $0.6$~ms per decision versus roughly $370$-$550$~ms for AO-WMMSE solvers.
\end{abstract}


\glsresetall

\section{Introduction}
The demand for high connectivity continues to grow, driven by massive \gls{iot} connectivity. Wireless traffic is expected to increase dramatically toward 2030, posing significant challenges to the spectral and energy efficiency of existing \gls{5g} networks \cite{wei2023integrated}. This growth requires ultra-reliable low-latency communications, higher density, and more adaptive systems. \Gls{6g} systems address current limitations \cite{david20186g}.

Among the \gls{6g} enablers, \gls{ris} stands out as a low-power way to shape radio propagation \cite{salim2025cooperative}. An \gls{ris} is a thin surface with many passive elements \cite{bjornson2022spm}. By adjusting these elements, it can steer, reflect, or focus signals to improve coverage, reduce interference, and support beamforming. In parallel, \glspl{uav} offer fast and flexible 3D placement. They can be deployed to restore \gls{los} paths, bypass temporary blockages, and provide on-demand connectivity for IoT deployments \cite{cheng2023ai}. Mounting an \gls{ris} on a \gls{uav} enables programmable control of the radio environment to extend coverage and improve throughput.

Realizing these benefits in practice requires careful treatment of two sources of uncertainty that the literature often simplifies. The first is \gls{uav} jitter caused by wind and control errors, which alters the effective array geometry and degrades beam alignment \cite{yuan2020jittering}. Another source of uncertainty is imperfect \gls{csi}. Designs optimized under the assumption of perfect channel knowledge can suffer significant performance loss when the available channel estimates are outdated or corrupted by noise. Large numbers of low-cost \gls{iot} nodes and limited overhead make frequent high-quality \gls{csi} acquisition difficult \cite{zheng2022survey}. In addition, \gls{ris} elements are passive and do not take part in pilot processing, which further complicates estimation.

Classical model-based designs are effective when channel models and \gls{csi} are accurate. However, \gls{uav} attitude variations and cascaded \gls{csi} uncertainty increase sensitivity to mismatch and reduce scalability \cite{shamasundar2023channel}. Because the joint design is nonconvex, common methods such as \gls{ao} and \gls{sdr} are only guaranteed to converge to stationary points, which can be suboptimal and often require many iterations \cite{wu2021tutorial}. Heuristic searches can reduce modeling effort but often trade solution quality and robustness for speed \cite{zhou2023survey}. These limitations motivate a data-driven controller that can adapt online under partial models. We therefore adopt \gls{drl} algorithms such as \gls{ddpg} and \gls{td3} for joint beamforming and \gls{ris} phase design.

\section{Related Work}
\label{sec:related_work}

The integration of learning-based optimization into wireless networks has evolved rapidly, moving from static terrestrial setups to mobility-aware aerial networks. First, we discuss foundational \gls{drl} approaches for static \gls{ris} deployments. Second, we examine the transition to \gls{uav}-mounted \gls{ris} architectures where mobility is a primary variable. Finally, we analyze existing robust designs, highlighting the gap in handling \gls{uav} jitter and imperfect \gls{csi}.

\subsection{Static RIS Optimization via DRL}

Early work applied \gls{drl} to jointly optimize beamforming and \gls{ris} phases. The authors in \cite{9110869} pioneered the use of \gls{drl} in this domain and showed that actor-critic algorithms like \gls{ddpg} could effectively maximize sum-rate (aggregate throughput) in continuous action spaces. However, a critical evolution in this literature questions the necessity of full \gls{mdp} formulations for wireless channels. Recognizing that channel realizations in block-fading models are often \gls{iid}, recent studies have advocated for simplifying the problem to a contextual bandit formulation. For instance, the study in \cite{9838369} demonstrated that a bandit-based approach could give comparable \gls{drl} performance with significantly lower computational overhead by discarding the temporal dependency of states. Similarly, the work in \cite{10054092} proposed a deep contextual bandit \gls{ddpg} framework, arguing that the bandit approach is more consistent with the i.i.d. channel model and offers sample-efficient alignment with the physical system.

\subsection{UAV-Mounted RIS Networks}
To overcome the blockage limitations of terrestrial deployments, recent research has shifted toward integrating \glspl{ris} with \glspl{uav}. To address this complexity, recent works have adopted \gls{drl} to handle the high-dimensional state space more efficiently. For instance, the study in \cite{9919620} formulated the capacity maximization problem as an \gls{mdp}. They employed a \gls{ddqn} to jointly optimize the \gls{uav} heading and \gls{ris} phase shifts. The authors in \cite{10817512} proposed an \gls{ris}-equipped \gls{uav} \gls{rsma} framework where \gls{sca}/\gls{sdr} and a \gls{ddpg} agent jointly optimize \gls{uav} position, \gls{ris} phase shifts, and power allocation to maximize the sum secrecy rate. The authors in \cite{10812045} investigated a \gls{uav}-mounted \gls{ris} \gls{swipt} system. They adopted a \gls{td3}-based \gls{drl} framework to jointly optimize the time-space-splitting factor, \gls{bs} transmit power, and \gls{ris} coefficients so as to maximize \gls{eh} efficiency. Despite their algorithmic differences, these studies assumed stable UAV and ideal or perfectly known CSI.

\begin{table*}[t]
\caption{Comparison of RIS-related works}
\centering
\renewcommand{\arraystretch}{1.5}
\resizebox{\textwidth}{!}{%
\begin{tabular}{cccccccc}
\hline
\textbf{Paper} & \textbf{Scenario} & \textbf{Metric} & \textbf{Variables} &
\textbf{Algorithm} & \textbf{Formulation} & \textbf{Imperfect CSI} &
\textbf{Jitter} \\
\hline
\cite{9110869}        & RIS              & Sum-rate                         & Beamformer + RIS 
                     & DDPG             & MDP        & \xmark                   & \xmark                 \\
\hline
\cite{9838369}        & Multi-RIS        & Sum-rate                         & Beamformer + RIS 
                     & Deep Contextual Bandit & CB    & \xmark                   & \xmark                 \\
\hline
\cite{10054092}       & RIS              & Sum-rate                         & Beamformer + RIS 
                     & DDPG             & CB         & \xmark                   & \xmark                 \\
\hline
\cite{9919620}        & UAV + RIS        & Capacity                         & UAV traj. + RIS 
                     & DDQN             & MDP        & \xmark                   & \xmark                 \\
\hline
\cite{10817512}       & UAV + RIS (RSMA) & Sum secrecy rate                 & Beamformer + RIS + UAV traj. 
                     & SCA + SDR + DDPG + AO & MDP   & \xmark                   & \xmark                 \\
\hline
\cite{10812045}  & UAV + RIS        & EH efficiency                   & TS factor $\tau$ + Power + RIS 
                     & TD3              & MDP        & \xmark                   & \xmark                 \\
\hline                     
\cite{9293148}        & RIS              & Secrecy / power                  & Beamformer + RIS 
                     & AO + SDR   & ---        & Statistical error         & \xmark                 \\
\hline
\cite{salim2025robust}  & UAV + RIS        & EH efficiency                   & TS factor $\tau$ + Beamformer + RIS 
                     & SD3              & MDP        & \xmark                         & Rotational-3D          \\
\hline
\cite{saglam2023deep} & RIS              & Sum-rate                         & Beamformer + RIS 
                     & SAC              & MDP        & Statistical error         & \xmark                 \\
\hline
\cite{adam2023intelligent}  & UAV (BS) + RIS (fixed) & Power & Beamformer + RIS  + UAV traj.  & AO + SDP + HSUDNN  & ---  & \xmark  & Bounded angular \\
\hline
Proposed              & UAV + RIS        & Throughput                         & Beamformer + RIS 
                     & DDPG + TD3       & CB         & Correlated statistical error        & Rotational-3D          \\
\hline
\end{tabular}
}
\label{tab:comparison}
\end{table*}

\subsection{Robustness to Imperfect CSI and UAV Jitter}

While the integration of \gls{uav} and \gls{ris} is promising, ensuring robustness against environmental uncertainties remains challenging. In terrestrial settings, the authors in \cite{9293148} studied robust secure transmission with imperfect cascaded \gls{csi}, jointly optimizing transmit beamforming and \gls{ris} phases via \gls{ao} with \gls{sdr}. The authors in \cite{saglam2023deep} considered an \gls{ris}-aided downlink and maximized the throughput using a \gls{sac} policy under hardware impairments and imperfect \gls{csi}. However, the scenario is terrestrial and does not model \gls{uav} jitter. 
The authors in \cite{salim2025robust} optimized \gls{eh} efficiency in a \gls{uav}-mounted \gls{ris} setting with three-dimensional rotational jitter using a smoothed \gls{sd3} \gls{drl} controller while imperfect \gls{csi} is not explicitly modeled. Collectively, these efforts treated a single impairment class, either \gls{csi}/hardware mismatch \cite{9293148,saglam2023deep,bansal2023ris} or \gls{uav} jitter \cite{salim2025robust}.

A closely related work considered mobility and impairments, but under different modeling choices. The authors in \cite{adam2023intelligent} studied a downlink system where the UAV acts as an aerial \gls{bs} and the \gls{ris} is fixed in space. They minimized transmit power by jointly optimizing the \gls{uav} trajectory, active beamforming, and \gls{ris} phases. Their robustness model relies on bounded angular perturbations for the UAV links and includes hardware impairments through a \gls{ris} phase noise model. These assumptions do not represent a UAV-mounted \gls{ris}, where three-dimensional rotational motion perturbs the \gls{ris} array geometry and directly impacts the cascaded channel. In addition, their uncertainty treatment does not model imperfect cascaded \gls{csi} as channel estimation uncertainty, since the phase-noise term captures hardware coefficient errors rather than cascaded-\gls{csi} quality. They further use a \gls{hsudnn} surrogate trained on outputs of an \gls{ao}/\gls{sdp}-based optimizer, which requires solver-generated supervision and differs from our model-free policy learning.

\subsection{Challenges and Motivation}
Table \ref{tab:comparison} summarizes recent \gls{ris}-assisted works and contrasts their problem settings and solution approaches. In the table, CB denotes contextual bandit, Jitter indicates whether \gls{uav} platform instability is explicitly modeled, and Imperfect \gls{csi} refers to channel estimation uncertainty. Most prior studies focus on static \gls{ris} deployments or idealized aerial links that assume perfect stability and channel knowledge, or they treat jitter and channel uncertainty separately. This leaves the combined impact of three-dimensional rotational jitter and imperfect cascaded \gls{csi} largely unaddressed in UAV-mounted \gls{ris} downlinks. Moreover, many baselines rely on iterative optimization such as \gls{ao}, which can incur high per-decision latency and limits rapid reconfiguration. This is particularly restrictive in dense \gls{iot} deployments, where pilot overhead and edge compute are limited and the controller must update quickly under time-varying links.

Motivated by the above, we present a practical \gls{miso} system where a ground \gls{bs} communicates with multiple users via a jitter-prone \gls{uav}-mounted \gls{ris}. We use a model-free \gls{drl} framework cast as a contextual bandit for optimization. We consider practical constraints such as 3D rotational jitter, channel estimation errors, a \gls{bs} transmit power constraint, and strict unit-modulus constraints on the \gls{ris} elements, providing a more resilient and scalable \gls{uav}-\gls{ris} architecture.
Specifically, the main contributions of this paper can be summarized as follows:
\begin{itemize}
\item A unified uncertainty model that captures \gls{uav} jitter and imperfect cascaded \gls{csi} is developed. This aligns the design with practical deployment impairments in \gls{uav}-mounted \gls{ris} links.
\item A constrained contextual bandit based \gls{drl} framework is proposed for robust joint \gls{bs} beamforming and \gls{ris} phase control under the developed uncertainty model. The proposed framework is instantiated with \gls{ddpg} and \gls{td3} backbones.
\item A Monte Carlo reward is introduced to estimate the expected throughput by resampling jitter and \gls{csi} error realizations at each decision step. A differentiable safety layer is introduced to project unconstrained policy outputs to satisfy the \gls{bs} transmit power constraint and the unit-modulus constraints on the \gls{ris} elements.
\item Comprehensive evaluation is conducted against \gls{ao} and \gls{wmmse}-based baselines, including a robust \gls{saa} variant and beamformer-only ablations. The results are complemented by a complexity analysis demonstrating sub-millisecond online inference and lower computational complexity than the baselines.
\end{itemize}
\section{System Model and Problem Formulation}
\label{sec:system_model}
\subsection{System Architecture}
\begin{figure}[t]
\centering
\includegraphics[width=0.9\linewidth]{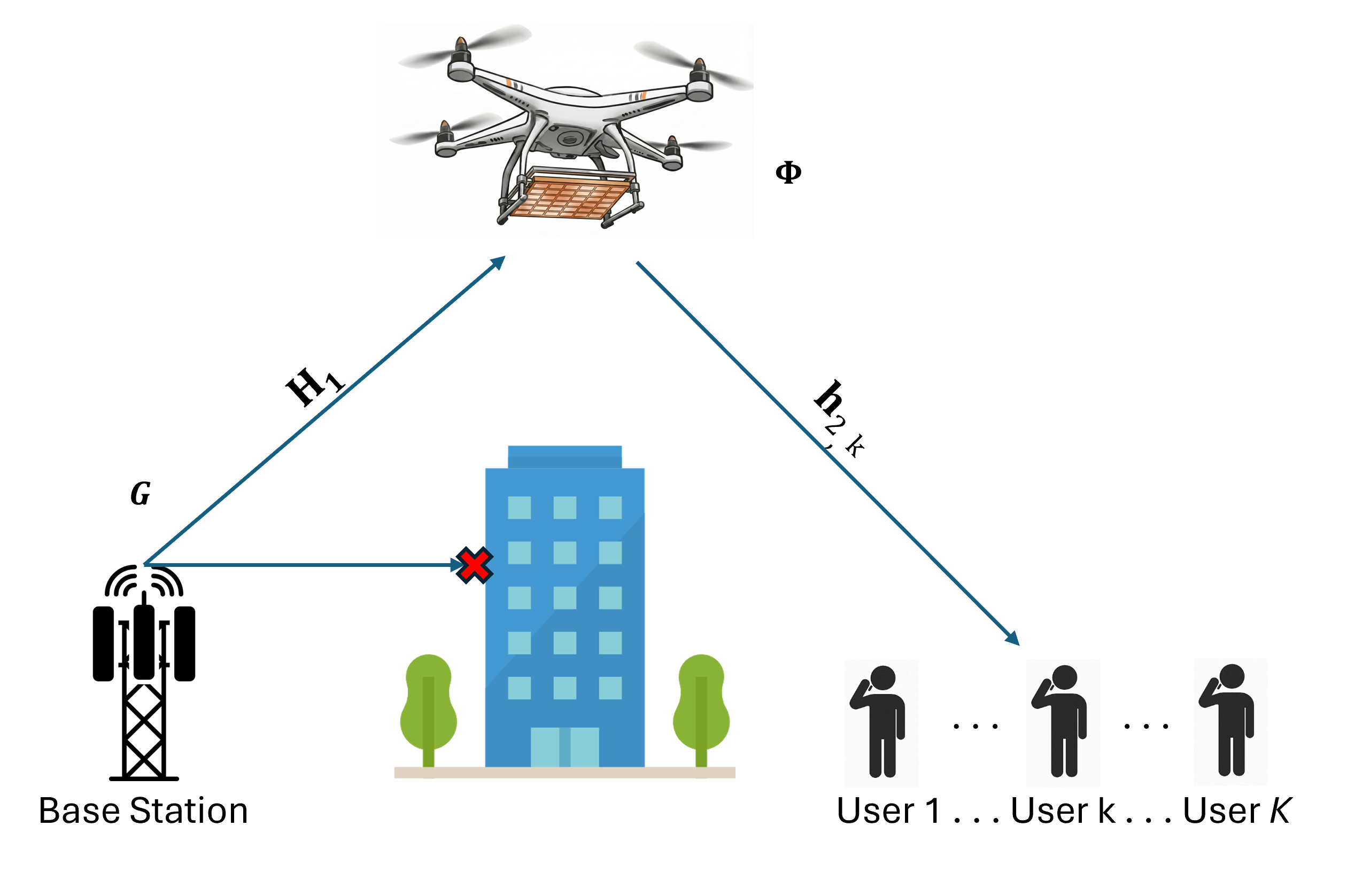}
\caption{\gls{uav}-mounted \gls{ris} communication system model}
\label{fig:1}
\end{figure}

We consider a downlink multiuser \gls{miso} communication system. We assume that the direct communication links between a ground-based \gls{bs} and multiple ground users are nonexistent due to severe blockages. To overcome this, a \gls{uav}-mounted \gls{ris} is employed to create a reflective communications path. The system, as illustrated in Fig.~\ref{fig:1}, is modeled within a 3D Cartesian coordinate system.

The system's transmitter is a ground-based \gls{bs}, located at a fixed position $\mathbf{p}_{\mathrm{BS}} \in \mathbb{R}^3$. It is equipped with a \gls{ula} of $M$ antennas oriented along the \(z\)-axis, with an inter-element spacing of $\lambda_c/2$, and operates under a maximum transmit power constraint, $P_{\max}$. Furthermore, the system serves $K$ single-antenna ground users. The locations of these $K$ users, denoted by $\{ \mathbf{p}_k \in \mathbb{R}^3 \}_{k=1}^K$, are randomly selected from a larger, predefined set of possible locations to increase generality. The communications model relies exclusively on the two-hop \gls{bs}-\gls{ris}-user link. We consider a \gls{uav} deployed at a fixed, static location $\mathbf{q}_{\mathrm{UAV}} \in \mathbb{R}^3$. The \gls{uav} carries an \gls{ris} with $N$ passive reflecting elements arranged in a \gls{upa} on the \(xy\)-plane, with an inter-element spacing of $\lambda_c/2$. Each element can apply an independent phase shift to the incident signal.
\subsection{Communications Model}
\label{sec:comm_model}
Communication from the \gls{bs} to the $K$ users is facilitated by the cascaded \gls{bs}-\gls{ris}-user link. The \gls{bs} employs a linear precoding matrix $\mathbf{G} \in \mathbb{C}^{M \times K}$ to encode the data symbol vector $\mathbf{x} = [x_1, \dots, x_K]^T \in \mathbb{C}^{K \times 1}$. We write $\mathbf{G} = [\mathbf{g}_1,\ldots,\mathbf{g}_K]$ with $\mathbf{g}_k \in \mathbb{C}^{M \times 1}$. Also, we assume the transmitted symbols are uncorrelated, each with unit power per stream, i.e., $\mathbb{E}[\mathbf{x}\mathbf{x}^H] = \mathbf{I}_K$. The signal received at the $k$-th user is then given by
\begin{IEEEeqnarray}{rCl}
    y_k & = & \mathbf{h}_{2,k}^H \boldsymbol{\Phi} \mathbf{H}_1 \mathbf{G} \mathbf{x} + n_k,
    \label{eq:received_signal}
\end{IEEEeqnarray}
where $\mathbf{H}_1 \in \mathbb{C}^{N \times M}$ is the channel matrix for the first hop (\gls{bs}-\gls{ris}), $\mathbf{h}_{2,k} \in \mathbb{C}^{N \times 1}$ is the channel vector for the second hop (\gls{ris}-user $k$), $n_k \sim \mathcal{CN}(0, \sigma_n^2)$ is the additive white Gaussian noise, and $\boldsymbol{\Phi} = \operatorname{diag}(e^{j\theta_1}, \dots, e^{j\theta_N}) \in \mathbb{C}^{N \times N}$ represents the phase shifts applied by the \gls{ris}. The first hop channel, $\mathbf{H}_1$, is based on an \gls{atg} model \cite{al2014optimal} whose large-scale path loss depends on the elevation angle, $\theta_{\mathrm{el}}$. The probability of a \gls{los} connection is \(P_{\mathrm{LoS}}(\theta_{\mathrm{el}})=(1+a\,\exp(-b(\theta_{\mathrm{el}}-a)))^{-1}\)
where $\theta_{\mathrm{el}}$ is in degrees, and $a$ and $b$ are environmental parameters. The total path loss gain, $\beta_1$, combines \gls{los} and \gls{nlos} components as
\begin{IEEEeqnarray}{rCl}
    \beta_1 & = & \kappa_1 \left( P_{\mathrm{LoS}}(\theta_{\mathrm{el}}) + (1 - P_{\mathrm{LoS}}(\theta_{\mathrm{el}})) \eta_{\mathrm{NLoS}} \right) d_1^{-\alpha_1},
\end{IEEEeqnarray}
where $\kappa_1=1$ is the reference path loss gain, $\eta_{\mathrm{NLoS}}$ is the \gls{nlos} attenuation, $d_1 = \|\mathbf{q}_{\mathrm{UAV}} - \mathbf{p}_{\mathrm{BS}}\|_2$ is the \gls{bs}-\gls{ris} distance, and $\alpha_1$ is the path loss exponent. The small-scale fading is modeled as Rician, and the full channel matrix is a superposition of the deterministic \gls{los} and stochastic \gls{nlos} components:
\begin{IEEEeqnarray}{rCl}
    \mathbf{H}_1 & = & \beta_1^{1/2} \left( \left( K_1/(K_1+1) \right)^{1/2} \mathbf{H}_{1,\mathrm{LoS}} \right. \nonumber \\
    & & \left. + \left( 1/(K_1+1) \right)^{1/2} \mathbf{H}_{1,\mathrm{NLoS}} \right), \label{eq:H1}
\end{IEEEeqnarray}

\noindent where $K_1$ is the Rician K-factor for the \gls{bs}-\gls{ris} link, and $\mathbf{H}_{1,\mathrm{NLoS}}$ consists of i.i.d. $\mathcal{CN}(0,1)$ random variables. Let the matrices $\mathbf{P}_{\mathrm{BS}} \in \mathbb{R}^{M \times 3}$ and $\mathbf{P}_{\mathrm{RIS}} \in \mathbb{R}^{N \times 3}$ contain the Cartesian coordinates of the \gls{bs} antenna elements and \gls{ris} elements, respectively. The deterministic component, $\mathbf{H}_{1,\mathrm{LoS}}$, is derived from the array geometry as the outer product of the array steering vectors:
\begin{IEEEeqnarray}{rCl}
    \mathbf{H}_{1,\mathrm{LoS}} & = & \mathbf{a}_{\mathrm{RIS}}(\mathbf{P}_{\mathrm{RIS}}, \mathbf{u}_{\mathrm{AoA}}) \mathbf{a}_{\mathrm{BS}}^H(\mathbf{P}_{\mathrm{BS}}, \mathbf{u}_{\mathrm{AoD}}),
\end{IEEEeqnarray}
\noindent where $\mathbf{a}_{\mathrm{RIS}}(\cdot)\in\mathbb{C}^{N\times 1}$ and $\mathbf{a}_{\mathrm{BS}}(\cdot)\in\mathbb{C}^{M\times 1}$. The steering vector $\mathbf{a}(\mathbf{P}, \mathbf{u}) \in \mathbb{C}^{L \times 1}$ for an array of $L$ elements and a unit direction vector $\mathbf{u} \in \mathbb{R}^{3 \times 1}$ is defined such that its $l$-th element is
\begin{equation}
    [\mathbf{a}(\mathbf{P}, \mathbf{u})]_l = \exp\left(-j \frac{2\pi}{\lambda_c} \mathbf{p}_l^T \mathbf{u}\right),
    \label{eq:steering_vector_element}
\end{equation}
where $\mathbf{p}_l^T$ is the $l$-th row of $\mathbf{P}$. For the first hop, the unit direction vectors are $\mathbf{u}_{\mathrm{AoD}} = (\mathbf{q}_{\mathrm{UAV}} - \mathbf{p}_{\mathrm{BS}}) / d_1$ and $\mathbf{u}_{\mathrm{AoA}} = -\mathbf{u}_{\mathrm{AoD}}$.
\noindent The second hop channel, $\mathbf{h}_{2,k}\in\mathbb{C}^{N\times 1}$, models the link from the aerial \gls{ris} to ground user $k$. The large-scale path loss gain is defined as $\beta_{2,k} = \kappa_2 d_{2,k}^{-\alpha_2}$, where $\kappa_2 > 0$ is the reference path loss gain and $d_{2,k} = \|\mathbf{p}_k - \mathbf{q}_{\mathrm{UAV}}\|_2$. The small-scale fading is also Rician. Therefore, the \gls{ris}-user $k$ channel can be given as
\begin{IEEEeqnarray}{rCl}
    \mathbf{h}_{2,k} & = & \beta_{2,k}^{1/2} \left( \left( K_2/(K_2+1) \right)^{1/2} \mathbf{h}_{2,k,\mathrm{LoS}} \right. \nonumber \\
    & & \left. + \left( 1/(K_2+1) \right)^{1/2} \mathbf{h}_{2,k,\mathrm{NLoS}} \right), \label{eq:h2k}
\end{IEEEeqnarray}
where $K_2$ is the Rician K-factor for the \gls{ris}-user link, $\mathbf{h}_{2,k,\mathrm{LoS}}\in\mathbb{C}^{N\times 1}$, and $\mathbf{h}_{2,k,\mathrm{NLoS}}\in\mathbb{C}^{N\times 1}$ is a vector of i.i.d.\ $\mathcal{CN}(0,1)$ variables. The deterministic \gls{los} component, $\mathbf{h}_{2,k,\mathrm{LoS}}$, is the \gls{ris} array steering vector towards user $k$, given by $\mathbf{a}_{\mathrm{RIS}}(\mathbf{P}_{\mathrm{RIS}}, \mathbf{u}_{\mathrm{AoD},k})$, where $\mathbf{u}_{\mathrm{AoD},k}=(\mathbf{p}_k-\mathbf{q}_{\mathrm{UAV}})/d_{2,k}$.

To formulate the system objective, we first define the aggregate second hop channel matrix as $\mathbf{H}_2=[\mathbf{h}_{2,1},\dots,\mathbf{h}_{2,K}]\in\mathbb{C}^{N\times K}$. The overall effective channel from the \gls{bs} to the users is then $\mathbf{H}_{\mathrm{eff}}=\mathbf{H}_2^H\boldsymbol{\Phi}\mathbf{H}_1\in\mathbb{C}^{K\times M}$. The $k$-th row of this matrix, denoted $\mathbf{h}_{\mathrm{eff},k}^H$, represents the end-to-end channel for user $k$. With the precoding matrix columns defined as $\mathbf{G}=[\mathbf{g}_1,\dots,\mathbf{g}_K]$, the \gls{sinr} for user $k$ is
\begin{IEEEeqnarray}{rCl}
    \operatorname{SINR}_k & = & \frac{\left| \mathbf{h}_{\mathrm{eff},k}^H \mathbf{g}_k \right|^2}{\sum_{j=1,\, j\neq k}^{K} \left| \mathbf{h}_{\mathrm{eff},k}^H \mathbf{g}_j \right|^2 + \sigma_n^2}.
    \label{eq:sinr}
\end{IEEEeqnarray}
\subsection{Uncertainty and Jitter Models}
\label{sec:uncertainty_jitter}
To capture the challenges of a practical deployment, we depart from the ideal assumption of perfect system knowledge and distinguish three cascaded channel representations. The estimated channel, $\mathbf{H}^{\mathrm{est}}$, denotes the cascaded channel estimate available to the controller. In our formulation, it corresponds to a nominal, non-jittered channel and does not include an explicit \gls{csi} error term. The jittered channel, $\tilde{\mathbf{H}}$, is an intermediate, unobserved channel obtained by perturbing the nominal propagation geometry with \gls{uav} rotational jitter. The true physical channel, $\mathbf{H}^{\mathrm{true}}$, is the channel used for throughput evaluation and is obtained by further corrupting the jittered channel with a cascaded \gls{csi} error. Thus, the controller observes $\mathbf{H}^{\mathrm{est}}$, while the throughput is evaluated using $\mathbf{H}^{\mathrm{true}}$.
The first source of uncertainty, \gls{uav} jitter, is modeled as a rotational instability. We consider small angle deviations, allowing us to focus on phase alignment changes rather than gain pattern variations. Specifically, $\delta_x, \delta_y, \delta_z \sim \mathcal{N}(0,\sigma_j^2)$ in radians around the roll, pitch, and yaw axes, respectively, with 
$\sigma_j \leq 0.175$~rad (corresponding to $10^{\circ}$). These induce a composite 3D rotation matrix {\footnotesize $\mathbf{R}_{\mathrm{jitt}} = \mathbf{R}_{\text{yaw}}(\delta_z)\,
\mathbf{R}_{\text{pitch}}(\delta_y)\, \mathbf{R}_{\text{roll}}(\delta_x)$}, where

\vspace{-0.05in}
{\footnotesize
\begin{align}
\mathbf{R}_{\text{yaw}}(\delta_z) &=
\begin{bmatrix}
\cos\delta_z & -\sin\delta_z & 0 \\
\sin\delta_z &  \cos\delta_z & 0 \\
0            &  0            & 1
\end{bmatrix}, \\
\mathbf{R}_{\text{pitch}}(\delta_y) &=
\begin{bmatrix}
 \cos\delta_y & 0 & \sin\delta_y \\
 0            & 1 & 0           \\
-\sin\delta_y & 0 & \cos\delta_y
\end{bmatrix}, \\
\mathbf{R}_{\text{roll}}(\delta_x) &=
\begin{bmatrix}
1 & 0           & 0            \\
0 & \cos\delta_x & -\sin\delta_x \\
0 & \sin\delta_x &  \cos\delta_x
\end{bmatrix}.
\end{align}
}
This rotation perturbs the \gls{ris} element coordinates:
\begin{IEEEeqnarray}{rCl}
    \tilde{\mathbf{P}}_{\mathrm{RIS}} & = & \mathbf{P}_{\mathrm{RIS}} \mathbf{R}_{\mathrm{jitt}}^{\mathsf T}.
\end{IEEEeqnarray}
The rotated coordinate matrix $\tilde{\mathbf{P}}_{\mathrm{RIS}}$ is then used to calculate the jittered \gls{los} components, $\tilde{\mathbf{H}}_{1,\mathrm{LoS}}$ and $\tilde{\mathbf{h}}_{2,k,\mathrm{LoS}}$.
We form $\tilde{\mathbf{H}}_{1}$ and $\tilde{\mathbf{h}}_{2,k}$ by combining the jittered \gls{los} components with the unchanged \gls{nlos} components.
For user $k$, the jittered cascaded channel is defined as
\begin{IEEEeqnarray}{rCl}
\tilde{\mathbf{H}}_{\mathrm{cascaded},k}
& \triangleq & \operatorname{diag}\!\big(\tilde{\mathbf{h}}_{2,k}^{H}\big)\,\tilde{\mathbf{H}}_{1}
\in \mathbb{C}^{N \times M}.
\end{IEEEeqnarray}
This represents the physical links under jitter while still assuming perfect \gls{csi}.
The second source of uncertainty is imperfect \gls{csi}. We adopt a practical cascaded error model, in which the true physical channel $\mathbf{H}_{\mathrm{cascaded},k}^{\mathrm{true}}$ is a noisy version of the jittered channel $\tilde{\mathbf{H}}_{\mathrm{cascaded},k}$ and is modeled as
\begin{IEEEeqnarray}{rCl}
\label{eq:cascaded_error_model}
    \mathbf{H}_{\mathrm{cascaded},k}^{\mathrm{true}} & = & \rho\,\tilde{\mathbf{H}}_{\mathrm{cascaded},k} + \sqrt{1-\rho^2}\,\mathbf{E}_k,
\end{IEEEeqnarray}
where $\rho \in [0,1]$ is the \gls{csi} quality parameter. The term $\mathbf{E}_k \in \mathbb{C}^{N \times M}$ is a random error matrix with i.i.d.\ entries $\mathcal{CN}(0,\sigma_k^2)$, independent across users and of the jitter variables and small-scale fading, with variance $\sigma_k^2=\beta_1\beta_{2,k}$.
This ensures the error power is appropriately scaled relative to the expected channel power for each user.

\subsection{Problem Formulation}
\label{sec:problem_formulation}

We formulate the problem of robust joint beamforming and \gls{ris} phase design. The objective is to design the \gls{bs} beamforming matrix $\mathbf{G}$ and the \gls{ris} phase shift matrix $\boldsymbol{\Phi}$ based only on the available channel estimates.

The instantaneous throughput is defined as
\begin{equation}
    T
    \triangleq \sum_{k=1}^{K} \log_2\bigl(1 + \operatorname{SINR}_k^{\mathrm{true}}\bigr),
    \label{eq:throughput_def}
\end{equation}
where $\operatorname{SINR}_k^{\mathrm{true}}$ denotes the \gls{sinr} for user $k$ computed using the true physical cascaded channel $\mathbf{H}_{\mathrm{cascaded}, k}^{\mathrm{true}}$.

The optimization problem is then
\begin{subequations}
\begin{IEEEeqnarray}{l.s.l}
    \underset{\mathbf{G}, \boldsymbol{\Phi}}{\text{maximize}} & \quad &
    \mathbb{E}\!\left[ T \right]
    \IEEEyessubnumber \label{eq:objective} \\
    \text{s.t.} & \quad & \operatorname{tr}(\mathbf{G}\mathbf{G}^H) \le P_{\max}, \IEEEyessubnumber \label{eq:power_constraint} \\
    & & |[\boldsymbol{\Phi}]_{n,n}| = 1,\; \forall n. \IEEEyessubnumber \label{eq:ris_constraint}
\end{IEEEeqnarray}
\end{subequations}
The expectation in \eqref{eq:objective} is taken over all random quantities in the channel model, including user locations, small-scale fading, \gls{uav} jitter, and cascaded \gls{csi} errors. Constraint~\eqref{eq:power_constraint} limits the \gls{bs} transmit power to $P_{\max}$, and \eqref{eq:ris_constraint} enforces strict unit-modulus coefficients on all \gls{ris} elements. Solving \eqref{eq:objective} is challenging because the objective is nonconvex, the expectation does not admit a closed-form expression, and the unit-modulus constraint further complicates the feasible set.
\section{Proposed Deep Reinforcement Learning Framework}
\subsection{DRL Overview and Formulation}
\label{ssec:drl_overview}
We adopt an actor-critic \gls{drl} architecture to handle the continuous actions in the joint beamforming and \gls{ris} phase design. The actor represents a deterministic policy that maps a state $s$ to an action $a$. The critic learns the action value function $Q(s,a)$ and is trained using the Bellman relation
\begin{IEEEeqnarray}{rCl}
Q(s,a) & = & r(s,a)+\gamma Q(s^{\prime},a^{\prime}),
\label{eq:bellman}
\end{IEEEeqnarray}
where $r(s,a)$ is the immediate reward, $s^{\prime}$ is the next state, $a^{\prime}$ is the next action, and $\gamma\in[0,1)$ is the discount factor.

The standard framework for modeling sequential decision-making problems in \gls{drl} is the \gls{mdp}. A special case arises when an agent's actions do not influence future states, that is, $\mathcal{P}(s'|s,a) = \mathcal{P}(s'|s)$. This collapses the full \gls{mdp} into a contextual bandit problem \cite{lattimore2020bandit}. In this paradigm, at each step, the agent observes a context and must choose an action to maximize the immediate reward, without needing to plan for future consequences. This bandit setting is formally enforced in the \gls{drl} framework by running one-step episodes and by setting the discount factor $\gamma=0$. This reduces the Bellman equation to $Q(s, a) = r(s, a)$, compelling the agent to learn a myopic policy that is optimal for the current context only. Furthermore, specialized contextual bandit algorithms can be more sample-efficient than full \gls{drl} methods but often struggle to scale to problems with high-dimensional continuous action spaces.
\subsection{DRL Algorithms}
\label{sec:rl_algorithms}
Our implementation is based on \gls{ddpg} \cite{lillicrap2015continuous}, and its successor, \gls{td3} \cite{fujimoto2018addressing}. However, standard implementations of these algorithms, which typically use a $\tanh$ output layer, are insufficient for our problem constraints. Furthermore, in our contextual bandit setting, the immediate reward is $y = r(s,a)$ with no bootstrap term, so we do not maintain separate target networks for the actor and critic.

We introduce a custom-constrained architecture where the actor network, $\mu(\mathbf{s}\,|\,\theta^\mu)$, outputs a raw, unconstrained vector. This vector is processed by a \gls{dsl}, $\Pi(\cdot)$, which implements a differentiable projection (safety layer) to enforce both the power constraint \eqref{eq:power_constraint} and the unit-modulus constraint \eqref{eq:ris_constraint} \cite{dalal2018safety}. The final policy is thus $\pi(\mathbf{s})=\Pi(\mu(\mathbf{s}\,|\,\theta^\mu))$. Each critic is trained to minimize the mean squared error relative to the immediate reward, $y=r$. The loss for the critic is
\begin{IEEEeqnarray}{rCl}
    L(\theta^Q) & = & \mathbb{E}_{(\mathbf{s},\mathbf{a},r) \sim \mathcal{D}} \big[ \big(r - Q(\mathbf{s}, \mathbf{a}\,|\,\theta^Q)\big)^2 \big],
\end{IEEEeqnarray}
where $\mathcal{D}$ is a replay buffer. Because the \gls{dsl} is differentiable, gradients from the critic flow through the projection to the actor:
\begin{IEEEeqnarray}{rCl}
    \nabla_{\theta^\mu} J & \approx & \mathbb{E}_{\mathbf{s} \sim \mathcal{D}} \big[ \nabla_{\mathbf{a}} Q(\mathbf{s}, \mathbf{a}\,|\,\theta^Q) \big|_{\mathbf{a} = \pi(\mathbf{s})} \cdot \nabla_{\theta^\mu} \pi(\mathbf{s}) \big]. \label{eq:actor_gradient}
\end{IEEEeqnarray}
\noindent We add exploration noise to the raw actor output before projection so all exploratory actions remain feasible, where $\mathcal{N}\sim\mathcal{N}(\mathbf{0},\sigma^{2}\mathbf{I})$.

\begin{algorithm}[t]
\caption{Constrained Contextual Bandit DDPG Algorithm}
\label{alg:ddpg}
\begin{algorithmic}[1]
\STATE Initialize actor $\mu(\mathbf{s}\,|\,\theta^\mu)$, critic $Q(\mathbf{s},\mathbf{a}\,|\,\theta^Q)$.
\STATE Initialize replay buffer $\mathcal{D}$.
\FOR{each context (one-step episode)}
    \STATE Environment resets: a new context
    \STATE Receive the full state $\mathbf{s}$.
    \STATE Select action with noise-before-projection: $\mathbf{a} = \Pi(\mu(\mathbf{s}\,|\,\theta^\mu) + \mathcal{N})$.
    \STATE Execute action $\mathbf{a}$ and observe reward $r$.
    \STATE Store transition $(\mathbf{s}, \mathbf{a}, r)$ in replay buffer $\mathcal{D}$.
    \STATE Sample minibatch of size \( B \): \( \{(\mathbf{s}_j, \mathbf{a}_j, r_j)\}_{j=1}^B \) from \( \mathcal{D} \)
    \STATE Set critic target: $y_i = r_i$.
    \STATE Update the critic by minimizing $L(\theta^Q) = \frac{1}{B}\sum_i (y_i - Q(\mathbf{s}_i, \mathbf{a}_i\,|\,\theta^Q))^2$.
    \STATE Update the actor using the policy gradient \eqref{eq:actor_gradient}.
\ENDFOR
\end{algorithmic}
\end{algorithm}
While \gls{ddpg} provides a strong foundation, it can suffer from overestimated Q-values. To address this, we also implement \gls{td3} \cite{fujimoto2018addressing}. \gls{td3} introduces twin critics (via clipped double Q-learning) and delayed policy updates to stabilize training. In our contextual bandit formulation, the critic target simplifies to $y=r$, making the third \gls{td3} mechanism (target policy smoothing) unused. We therefore only utilize the twin critic and delayed update components to improve stability over \gls{ddpg}.

\subsection{State Space}
\label{ssec:state_space}

The state $\mathbf{s} \in \mathcal{S}$ provides the agent with the contextual information for the current instance. It concatenates the geometric distances with the nominal channel estimates. This is defined as:
\begin{IEEEeqnarray}{rCl}
\mathbf{s}
& = & \big(
\{ d_{2,k} \}_{k=1}^{K},
\, \Re\{\mathbf{H}_1^{\mathrm{est}}\},
\, \Im\{\mathbf{H}_1^{\mathrm{est}}\},
\nonumber \\
& & \quad \,
\Re\{\mathbf{H}_2^{\mathrm{est}}\},
\, \Im\{\mathbf{H}_2^{\mathrm{est}}\}
\big),
\end{IEEEeqnarray}
where $d_{2,k}$ is the distance between the \gls{ris} and user $k$, $\mathbf{H}_1^{\mathrm{est}}$ is the estimated \gls{bs}-\gls{ris} channel, and $\mathbf{H}_2^{\mathrm{est}} = [\mathbf{h}_{2,1}^{\mathrm{est}}, \ldots, \mathbf{h}_{2,K}^{\mathrm{est}}]$ forms the estimated \gls{ris}-user channels. The agent observes non-jittered, error-free channels and must learn a mapping from this state to actions that perform well on average under the unobserved jitter and cascaded \gls{csi} errors.

\subsection{Action Space}
\label{sec:action_space}
The action $\mathbf{a} \in \mathcal{A}$ is a continuous vector of dimension $2MK + 2N$ that encapsulates all the parameters the agent controls. This vector is defined as:
\begin{IEEEeqnarray}{rCl}
\mathbf{a}
& = & \big[
\operatorname{vec}(\Re\{\mathbf{G}\}), \,
\operatorname{vec}(\Im\{\mathbf{G}\}),
\nonumber \\
& & \quad \,
\Re\{\operatorname{diag}(\boldsymbol{\Phi})\}, \,
\Im\{\operatorname{diag}(\boldsymbol{\Phi})\}
\big].
\label{eq:action_vector}
\end{IEEEeqnarray}
The first $2MK$ entries of the action vector are reshaped into the complex \gls{bs} beamforming matrix $\mathbf{G} \in \mathbb{C}^{M \times K}$, using half of the entries for the real part and half for the imaginary part. This raw beamforming matrix is then passed through the \gls{dsl}, which enforces the transmit power constraint \eqref{eq:power_constraint} via a differentiable Frobenius-norm projection. In this way, the actor operates in an unconstrained Euclidean space, while the mapping to $\mathbf{G}$ always respects the \gls{bs} power budget.

The remaining $2N$ entries of the action vector are reshaped as $N$ two-dimensional real vectors, each corresponding to one \gls{ris} element. The \gls{dsl} normalizes each of these vectors to unit norm and maps them to complex coefficients on the diagonal of $\boldsymbol{\Phi}$. This normalization enforces the strict unit-modulus constraint in \eqref{eq:ris_constraint} for every \gls{ris} element, allowing the actor to output unconstrained values while ensuring that all executed actions are physically feasible.
\subsection{Reward Function}
\label{sec:reward_function}
The reward implements the objective in \eqref{eq:objective} using the instantaneous throughput defined in \eqref{eq:throughput_def}. It is evaluated on the true cascaded channels, which include \gls{uav} jitter and cascaded \gls{csi} errors, rather than on the nominal estimate observed by the agent. For a given action, the environment generates $S$ independent realizations of jitter and \gls{csi} errors, constructs the corresponding true channels, and computes the throughput $T^{(i)}$ for each realization as in \eqref{eq:throughput_def}. The reward is then defined as
\begin{equation}
    r = \frac{1}{S} \sum_{i=1}^{S} T^{(i)}.
    \label{eq:reward}
\end{equation}
This Monte Carlo averaging reduces the variance of the reward signal due to the unobserved jitter and \gls{csi} errors, making learning more stable while remaining consistent with the optimization objective.
\subsection{Training Dynamics}
\label{sec:training_dynamics}
The training process follows the contextual bandit formulation, where each episode corresponds to an independent problem instance. At the beginning of an episode, a \texttt{reset} generates a new context. We sample a new set of $K$ user locations from the predefined pool, recompute the corresponding large-scale and \gls{los} components of the \gls{ris}-user links, and regenerate the \gls{nlos} components of the 

Me\gls{bs}-\gls{ris} and \gls{ris}-user channels from their respective complex Gaussian distributions. We keep the \gls{bs}-\gls{ris} geometry and its large-scale and \gls{los} terms fixed. This yields a statistically independent channel scenario for each episode. Within an episode, the agent performs a single \texttt{step} and the episode terminates, implementing the one-step contextual bandit structure described earlier. Furthermore, this sampling scheme exposes the agent to diverse channel realizations consistent with the average throughput objective.
\section{Numerical Results}
\label{sec:results}
In this section, we evaluate the performance of our proposed \gls{drl} framework. We construct a custom simulation environment in Python using PyTorch to model the downlink \gls{miso} system described in Section \ref{sec:system_model}. All \gls{drl} agents (both \gls{ddpg} and \gls{td3}) are trained and evaluated on an NVIDIA GeForce RTX 3080. To ensure statistically significant and reproducible results, our evaluation methodology is aligned with the recommendations for rigorous \gls{drl} evaluation \cite{henderson2018deep}. All \gls{drl} agents are averaged over 10 distinct random seeds. The classical optimizers, however, are deterministic, require no seed averaging, and are plotted without confidence intervals. For convenience, jitter standard deviations are reported in degrees in the figures.
\noindent To validate our approach, we compare its performance against several key benchmark schemes. In the subsequent figures, these schemes are labeled as follows:
\begin{itemize}
    \item AO-WMMSE: A conventional iterative algorithm based on the \gls{wmmse} criterion, adapted with \gls{ao} to solve for the precoder and \gls{ris} phases separately \cite{shi2011iteratively}.
    \item AO-WMMSE-SAA: A robust, non-\gls{drl} benchmark that uses \gls{saa} with the AO-WMMSE algorithm. It optimizes under the same uncertainty distributions (jitter and imperfect \gls{csi}) as our \gls{drl} agent, serving as the primary traditional competitor for robust implementations \cite{Kim2011SAA}.
    \item TD3 (BF-Only) and DDPG (BF-Only): An ablation study of our proposed model. Here, the \gls{drl} agent is trained to optimize only the \gls{bs} beamforming matrix $\mathbf{G}$, while the \gls{ris} phase shifts $\boldsymbol{\Phi}$ are kept fixed to a random configuration. This baseline is used to quantify the performance gain achieved by actively optimizing the \gls{ris}.
\end{itemize}

\noindent The core idea of these experiments is to demonstrate the viability and robustness of our model-free \gls{drl} framework. We first establish the agents' convergence in an ideal environment. We then evaluate robustness to \gls{uav} jitter and imperfect \gls{csi}, considering these factors individually and jointly. Finally, we compare the computational complexity to highlight the practical deployment advantage of low online computational complexity.
\subsection{Hyperparameters}
\label{sec:hyperparameters}

The simulation environment models a realistic urban \gls{iot} scenario with a single \gls{bs} ($M=4$),
a \gls{uav}-mounted \gls{ris} ($N=16$), and $K=4$ users. The simulation, channel, and learning settings
are summarized in Table~\ref{tab:sim_learn_params}. Both the actor and critic are fully connected
\glspl{mlp} with two hidden layers of 256 units each and ReLU activations. We denote their total numbers of trainable parameters by $P_a$ and $P_c$, respectively.

\begin{table}[t]
\caption{Simulation Environment and Learning Parameters}
\label{tab:sim_learn_params}
\centering
\renewcommand{\arraystretch}{1.08}
\setlength{\tabcolsep}{5pt}
\footnotesize
\begin{tabular}{|l|l|}
\hline
\textbf{Parameter} & \textbf{Value} \\
\hline
\multicolumn{2}{|c|}{\textbf{System Parameters}} \\
\hline
BS antennas, $M$ & 4 \\
RIS elements, $N$ & 16 \\
Number of users, $K$ & 4 \\
Carrier frequency & 24 GHz \\
BS position, $\mathbf{p}_{\mathrm{BS}}$ & [0, 0, 20] m \\
UAV position, $\mathbf{q}_{\mathrm{UAV}}$ & [50, 0, 50] m \\
Max BS power, $P_{\max}$ & 1.0 W \\
Noise power, $\sigma_n^2$ & -131 dBW \\
\hline
\multicolumn{2}{|c|}{\textbf{Channel Parameters}} \\
\hline
ATG env. param., $a$ & 9.61 \\
ATG env. param., $b$ & 0.16 \\
Path-loss exp. (BS--RIS), $\alpha_1$ & 2.2 \\
NLoS attenuation, $\eta_{\mathrm{NLoS}}$ & 0.5 \\
Path-loss const. (RIS--user), $\kappa_2$ & 0.001 \\
Path-loss exp. (RIS--user), $\alpha_2$ & 2.4 \\
K-factor (BS--RIS), $K_1$ & 3 \\
K-factor (RIS--user), $K_2$ & 2 \\
Monte Carlo samples, $S$ & 6 \\
\hline
\multicolumn{2}{|c|}{\textbf{DRL Agent Hyperparameters}} \\
\hline
Learning rate & $1 \times 10^{-4}$ \\
Batch size, $B$ & 32 \\
Replay buffer size & 20{,}000 \\
Discount factor, $\gamma$ & 0.0 \\
Policy delay (TD3) & 3 \\
Training steps, $T$ & 200{,}000 \\
Evaluation episodes & 2{,}000 \\
RIS exploration noise (std.) & 0.1 \\
Beamformer exploration noise (std.) & 0.2 \\
\hline
\multicolumn{2}{|c|}{\textbf{AO-Based Baseline Settings}} \\
\hline
AO outer iterations, $a_{\max}$ & 70 \\
WMMSE inner iterations, $w_{\mathrm{in}}$ & 12 \\
Bisection steps, $n_{\mathrm{bisect}}$ & 30 \\
\hline
\end{tabular}
\end{table}

\begin{figure*}[!t]
    \centering
    \begin{minipage}{0.24\textwidth}
        \centering
        \includegraphics[width=\linewidth]{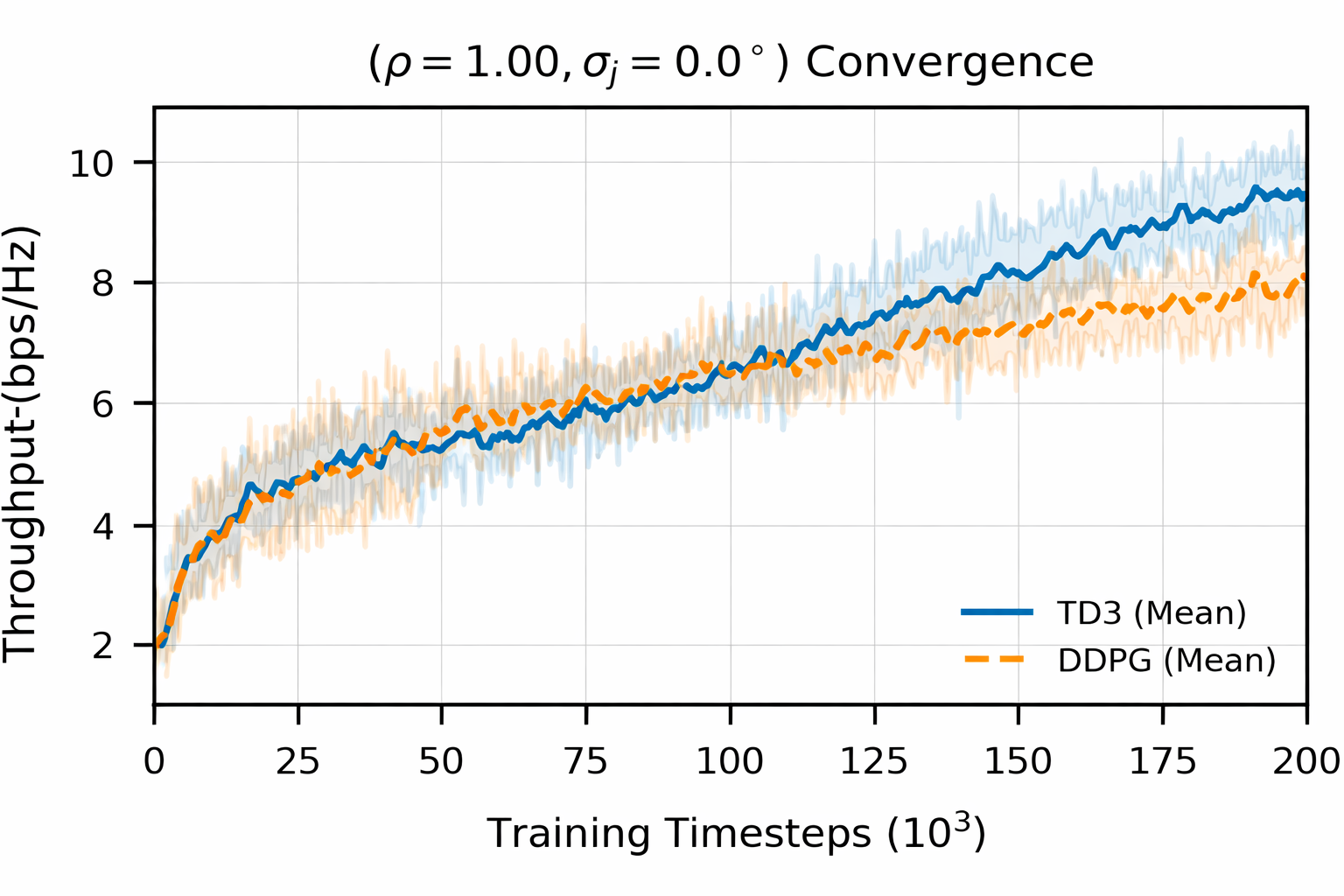}
        {\footnotesize \centerline{(a) Ideal}}
    \end{minipage}\hfill
    \begin{minipage}{0.24\textwidth}
        \centering
        \includegraphics[width=\linewidth]{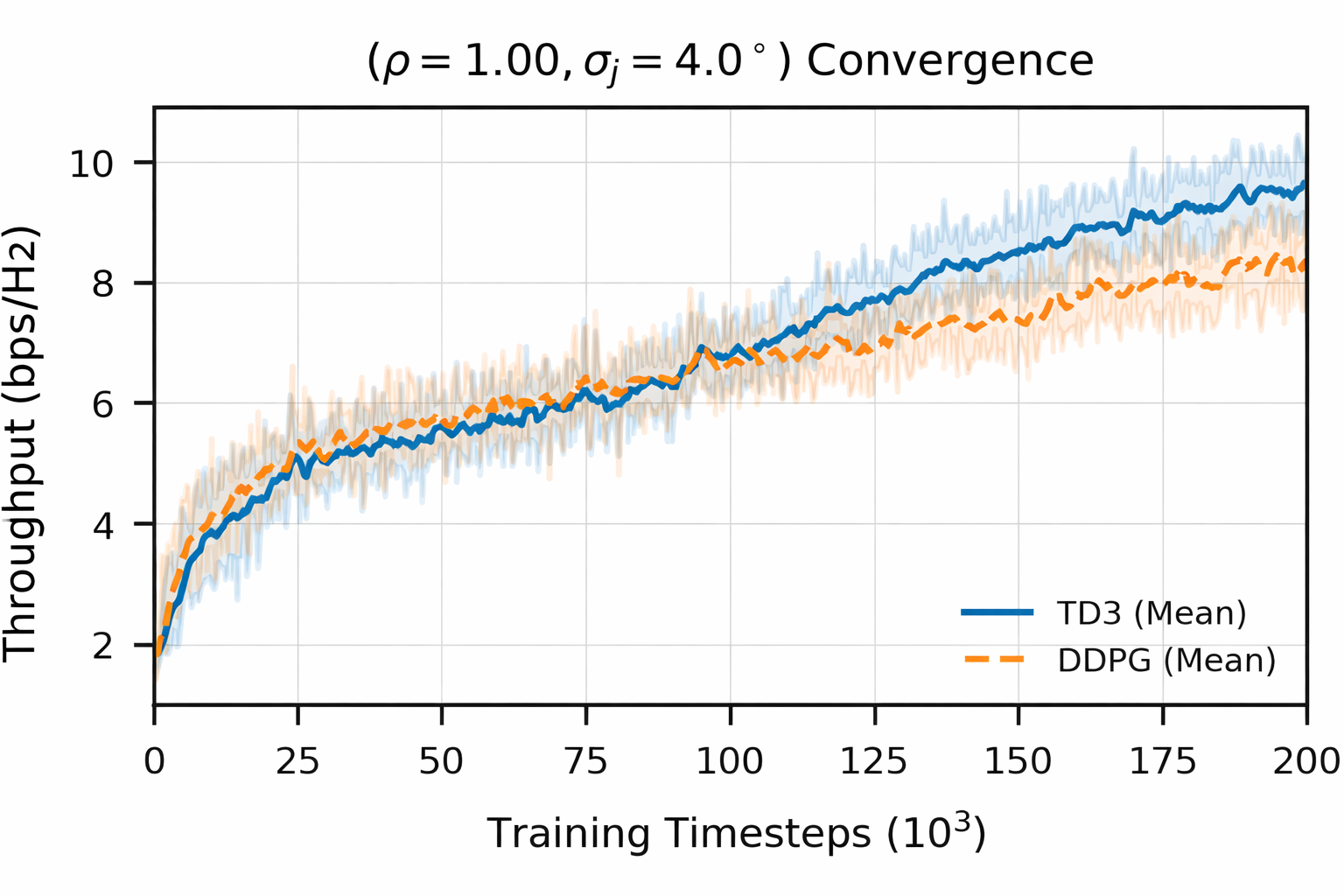}
        {\footnotesize \centerline{(b) Jitter}}
    \end{minipage}\hfill
    \begin{minipage}{0.24\textwidth}
        \centering
        \includegraphics[width=\linewidth]{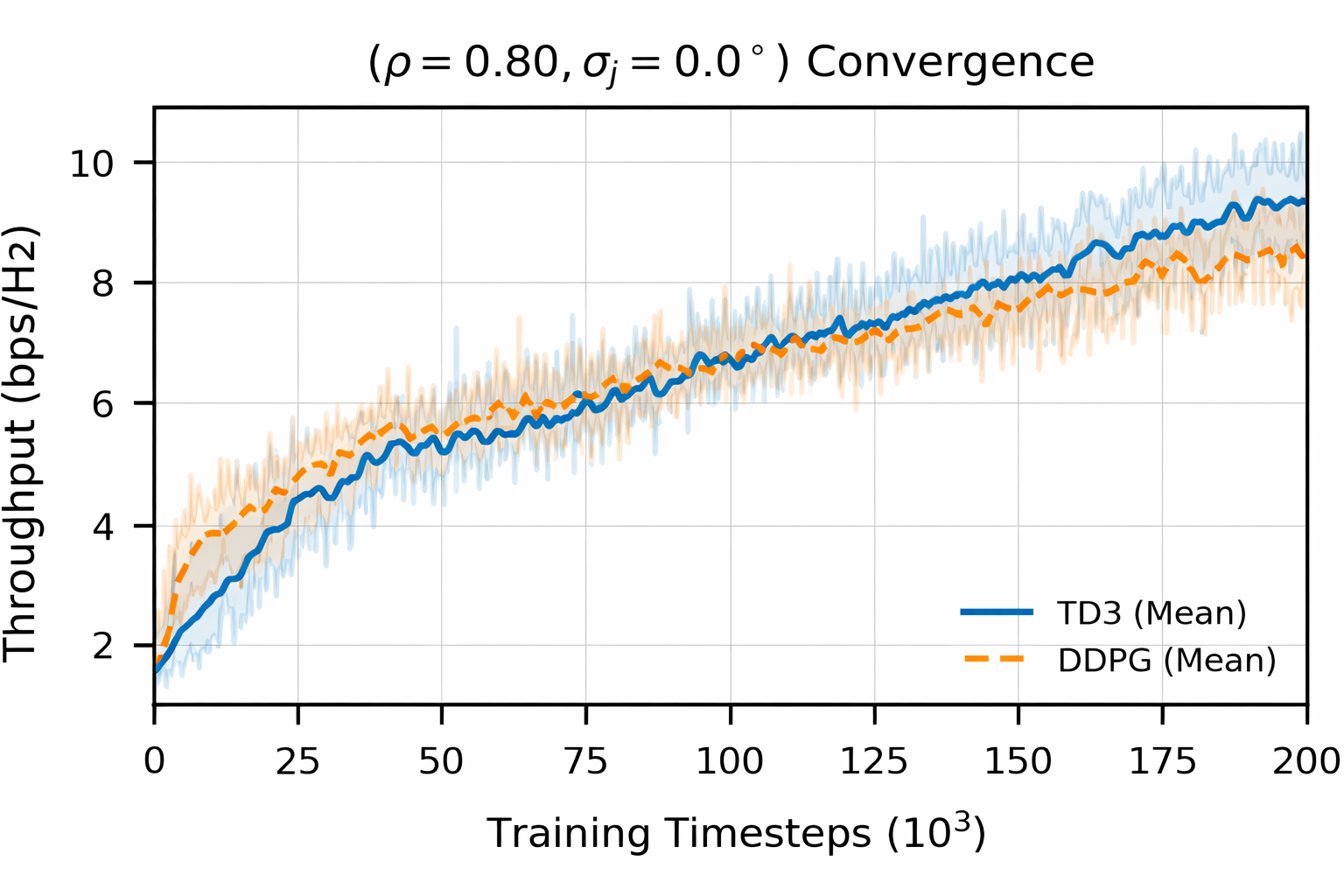}
        {\footnotesize \centerline{(c) Imperfect CSI}}
    \end{minipage}\hfill
    \begin{minipage}{0.24\textwidth}
        \centering
        \includegraphics[width=\linewidth]{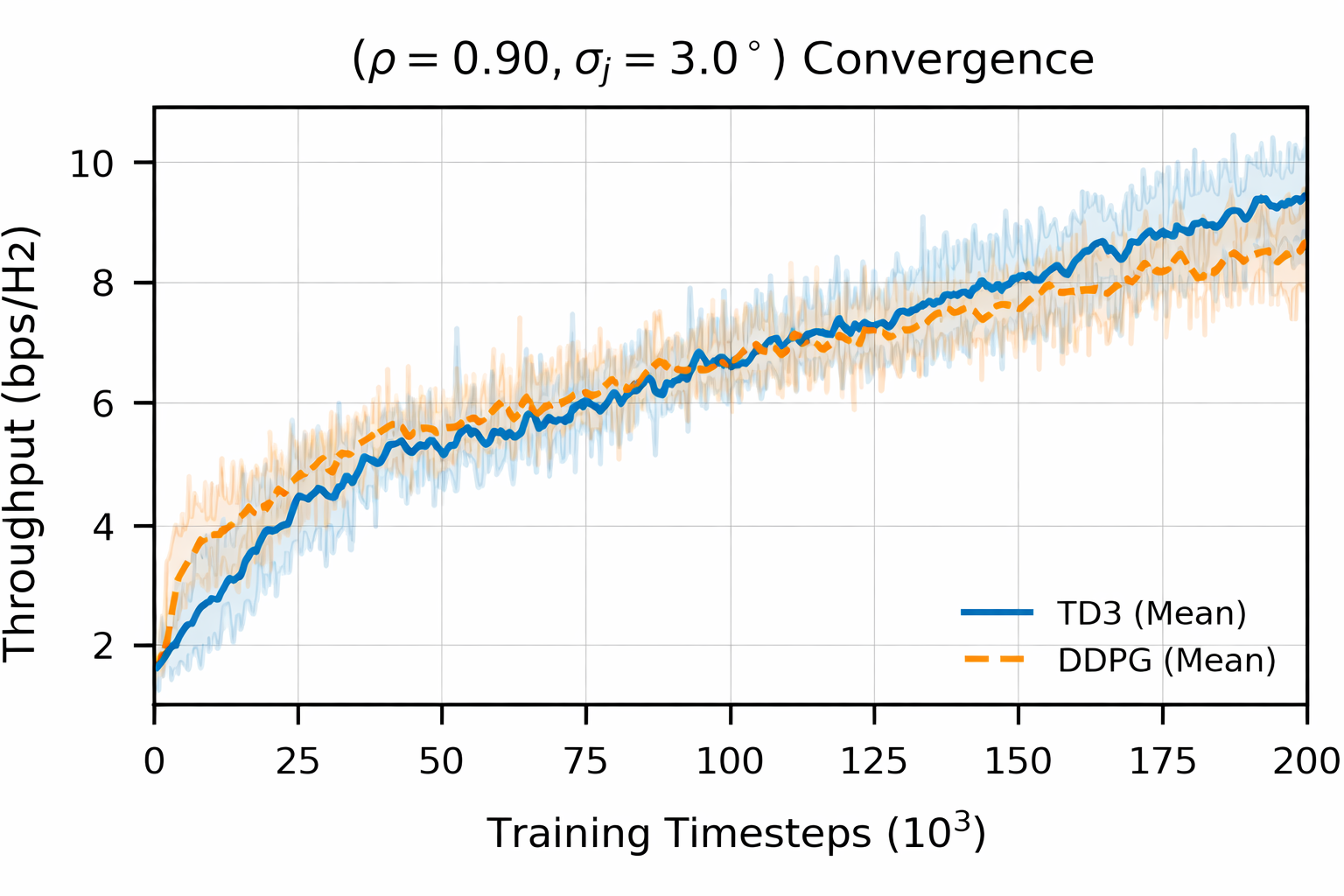}
        {\footnotesize \centerline{(d) Combined}}
    \end{minipage}
    \caption{Agent convergence for different scenarios.}
    \label{fig:training_convergence_combined}
\end{figure*}

\subsection{Convergence Analysis}
\label{ssec:convergence_analysis}
We first examine the convergence behavior of the \gls{ddpg} and \gls{td3} agents across four representative scenarios. Fig. 2(a) shows an ideal baseline with no jitter and perfect \gls{csi} ($\sigma_{j} = 0.0^\circ$, $\rho = 1.0$). Fig. 2(b) shows a jitter only case with \gls{uav} instability and perfect \gls{csi} ($\sigma_{j} = 4.0^\circ$, $\rho = 1.0$). Fig. 2(c) shows a \gls{csi} only case with imperfect estimation but a stable \gls{uav} ($\sigma_{j} = 0.0^\circ$, $\rho = 0.80$). Finally, Fig. 2(d) shows a joint uncertainty case with both jitter and \gls{csi} degradation ($\sigma_{j} = 3.0^\circ$, $\rho = 0.90$). For each scenario, both agents are trained using 10 independent training runs (random seeds) per algorithm. The x-axis shows training timesteps in thousands, the solid curves show the mean throughput (smoothed with a 2{,}000-step rolling window), and the shaded regions indicate 95\% confidence intervals across seeds computed from the unsmoothed data, as shown in Fig.~\ref{fig:training_convergence_combined}.

\noindent As illustrated in Fig.~\ref{fig:training_convergence_combined}, both \gls{ddpg} and \gls{td3} successfully learn in all environments, converging from a low initial throughput to a high, stable performance. This confirms that the \gls{drl} approach remains viable even when trained in stochastic environments with multiple uncertainties. The \gls{td3} algorithm, benefiting from its stability enhancements and twin critics that mitigate overestimation bias, consistently achieves a notably higher final throughput than \gls{ddpg}, with the improvement typically becoming apparent at around $10^5$ training steps for all cases. Under ideal conditions, as shown in Fig.~\ref{fig:training_convergence_combined}(a), the agents achieve a mean throughput of 11.08 bps/Hz for \gls{td3} and 9.18 bps/Hz for \gls{ddpg} in a separate 2{,}000-episode evaluation. Even under severe impairments, as in Fig.~\ref{fig:training_convergence_combined}(b)-(d), the agents do not diverge but instead learn robust policies adapted to each uncertainty setting, with \gls{td3} still outperforming \gls{ddpg}. Furthermore, the shaded regions are very narrow, indicating low variability across seeds and convergence of all runs to similar values with similar trajectories. This confirms that the training procedure is stable.

\subsection{Evaluation Analysis}
\label{ssec:evaluation_analysis}
We evaluate the final performance of the proposed \gls{drl} agents against the AO-WMMSE and robust AO-WMMSE-SAA baselines. We test under three types of uncertainty: \gls{uav} jitter with perfect \gls{csi}, imperfect \gls{csi} with no jitter, and joint jitter and \gls{csi} degradation. For each operating point (jitter level, \gls{csi} quality, or combined level), a separate agent is trained from scratch under the matched environment parameters. We train 10 such agents with different random seeds, and each trained agent is evaluated over 2{,}000 independent test episodes. Each point in the following plots represents the mean throughput over these evaluations, and the error bars indicate the $95\%$ confidence intervals computed across the 10 seeds. For the joint-uncertainty scenario, we define five combined error levels \(\mathrm{L}0\)–\(\mathrm{L}4\) as
\begin{equation}
\label{eq:combined_levels}
\left\{
\begin{aligned}
\mathrm{L}0 &: (\rho,\sigma_j)=(1.0,0^\circ), & \mathrm{L}1 &: (0.9,3^\circ),\\
\mathrm{L}2 &: (0.8,5^\circ), & \mathrm{L}3 &: (0.6,7^\circ),\\
\mathrm{L}4 &: (0.5,10^\circ).
\end{aligned}
\right.
\end{equation}

\begin{figure*}[!t]
    \centering
    \begin{minipage}{0.32\textwidth}
        \centering
        \includegraphics[width=\linewidth,height=0.7\linewidth]{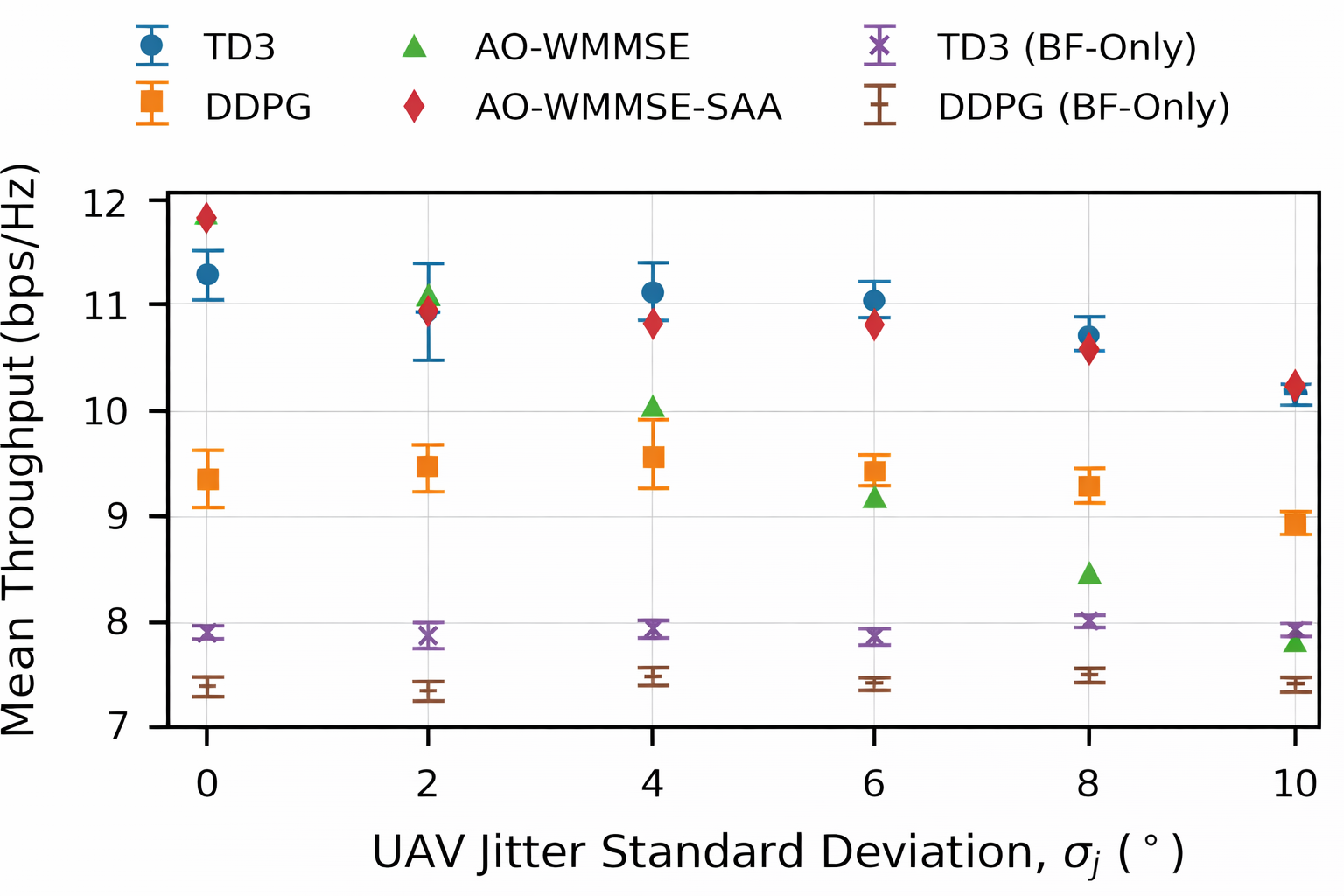}
        {\footnotesize \centerline{(a)}}
    \end{minipage}\hfill
    \begin{minipage}{0.32\textwidth}
        \centering
        \includegraphics[width=\linewidth,height=0.7\linewidth]{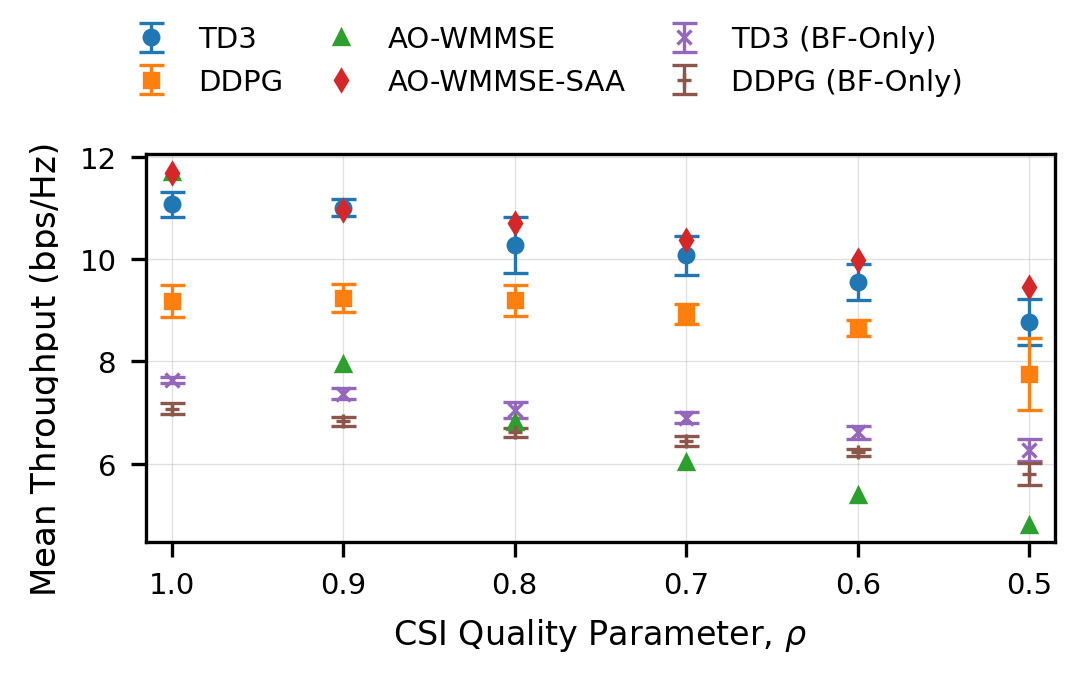}
        {\footnotesize \centerline{(b)}}
    \end{minipage}\hfill
    \begin{minipage}{0.32\textwidth}
        \centering
        \includegraphics[width=\linewidth,height=0.7\linewidth]{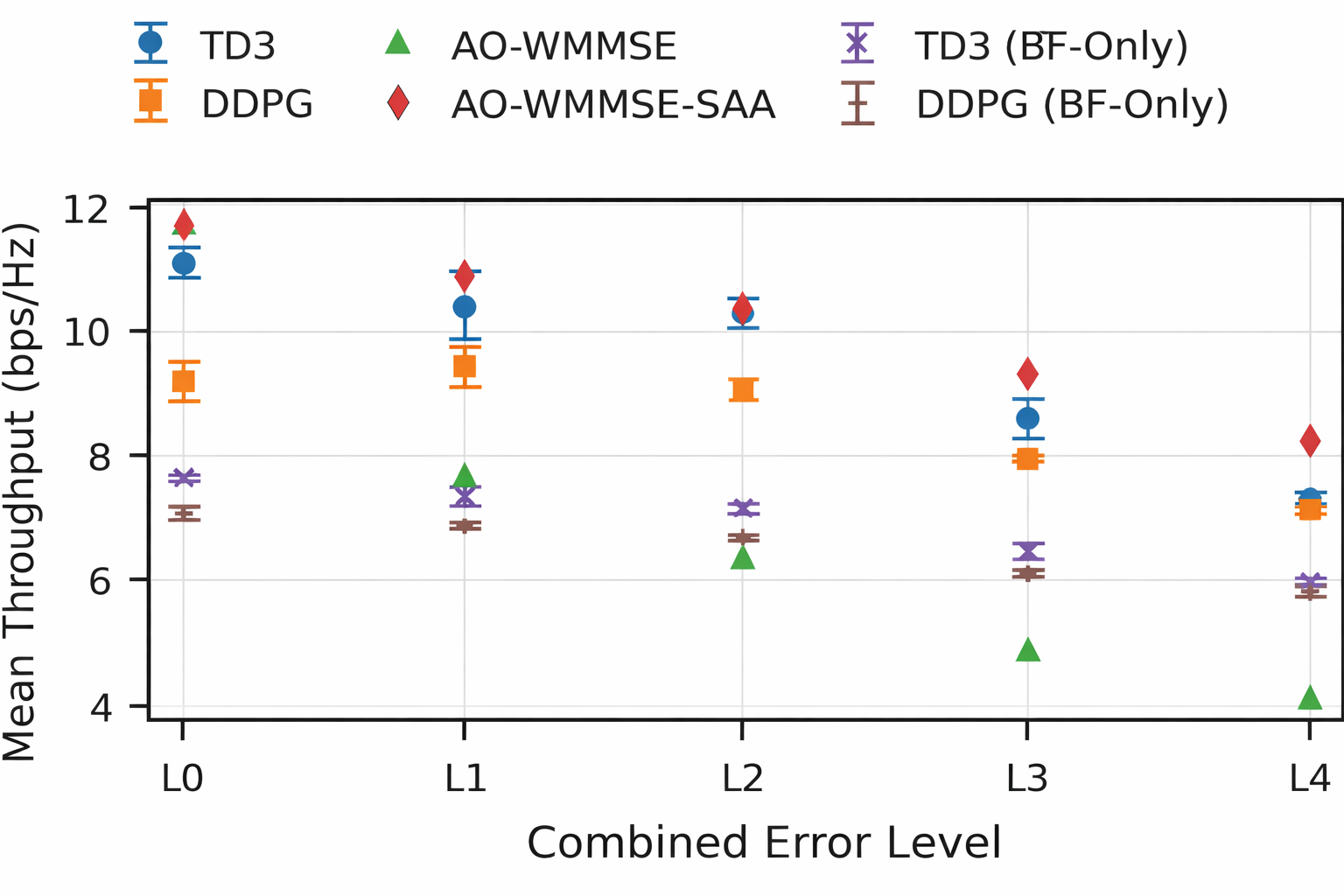}
        {\footnotesize \centerline{(c)}}
    \end{minipage}
    \caption{Evaluation figures: (a) Performance vs. UAV jitter ($\rho=1.0$), (b) Performance vs. CSI quality ($\sigma_j=0.0^\circ$), and (c) Performance vs. combined error level.}
    \label{fig:evaluation_figures}
\end{figure*}

Fig.~\ref{fig:evaluation_figures}(a) shows the impact of \gls{uav} jitter when \gls{csi} is perfect ($\rho = 1.0$). The conventional AO-WMMSE benchmark degrades by approximately 35.4\% as $\sigma_j$ increases from $0^\circ$ to $10^\circ$, indicating that a design optimized for a static geometry is highly sensitive to \gls{uav} jitter. In contrast, the \gls{drl} agents and the robust AO-WMMSE-SAA exhibit significantly milder degradation. The \gls{td3} agent’s throughput decreases by about 9.6\%, while the AO-WMMSE-SAA drops by 13.8\%. The \gls{ddpg} agent shows the smallest relative loss, with only 5.6\% degradation over the same jitter range.

Fig.~\ref{fig:evaluation_figures}(b) isolates the effect of \gls{csi} quality with a perfectly stable \gls{uav} ($\sigma_j = 0.0^\circ$). Here, AO-WMMSE, which assumes perfect \gls{csi}, suffers a severe performance loss of about 59\% as $\rho$ decreases from 1.0 to 0.5. The robust AO-WMMSE-SAA and the \gls{drl} agents are substantially more tolerant to \gls{csi} errors. Across all $\rho$ values, the \gls{td3} agent remains within roughly $\pm 7\%$ of the AO-WMMSE-SAA benchmark. The \gls{ddpg} agent operates $13$-$22\%$ below AO-WMMSE-SAA in absolute terms, but its relative degradation from $\rho=1.0$ to $0.5$ is only 15.5\%, comparable to the SAA benchmark’s 19.1\% drop.

Fig.~\ref{fig:evaluation_figures}(c) considers the combined uncertainty levels L0--L4 defined in~\eqref{eq:combined_levels}. As both jitter and \gls{csi} errors increase, AO-WMMSE’s performance collapses, with a total degradation of about 64.7\% from L0 to L4. The robust AO-WMMSE-SAA degrades by 29.4\%. The \gls{td3} agent remains within approximately 0-12\% of AO-WMMSE-SAA across all levels, while the \gls{ddpg} agent exhibits the smallest relative loss at 22.6\% but stays below SAA in absolute throughput. Together, these results confirm that the learned \gls{drl} policies maintain strong robustness under jitter-only, \gls{csi}-only, and combined uncertainty conditions.

The BF-only ablations across all three curves help separate the contributions of beamforming and \gls{ris} control to the overall performance. In the jitter robustness case, the mean throughput of TD3 and DDPG increase from $7.66$ and $7.13$~bps/Hz in the BF-only setting to $10.68$ and $9.15$~bps/Hz when the \gls{ris} is also optimized, corresponding to \gls{ris}-induced gains of approximately $3.02$ and $2.02$~bps/Hz (about $39\%$ and $28\%$ over their BF-only baselines). Under varying \gls{csi} quality, the gains are $3.15$ versus $2.33$~bps/Hz (about $45\%$ versus $36\%$), and in the combined-uncertainty case they are $2.60$ versus $2.03$~bps/Hz (about $38\%$ versus $31\%$). Across all scenarios, TD3’s beamformer alone is only modestly better than DDPG’s (by roughly $0.4$-$0.5$~bps/Hz), whereas the larger share of the performance gap comes from the additional gain obtained when the \gls{ris} is controlled. This indicates that TD3 improves not only the active beamforming but also the joint exploitation of the \gls{ris} degrees of freedom.
\subsection{Complexity Analysis}
\label{ssec:complexity_analysis}

Each method incurs the same environment cost, since every simulator step evaluates a Monte Carlo robust throughput with $S$ channel samples, requiring $\mathcal{O}(S K^2 N M)$ operations. On top of this, the DRL controllers differ only in their actor-critic update structure: DDPG performs one critic and one actor update per gradient step, leading to a training cost proportional to $B(P_c + P_a)$, i.e., $\mathcal{O}\big(B(P_c + P_a)\big)$, whereas TD3 maintains two critics and updates the actor once every $\delta$ critic updates, with per-step cost proportional to $B\big(2P_c + \tfrac{1}{\delta} P_a\big)$, which remains of the same order $\mathcal{O}\big(B(P_c + P_a)\big)$ but with different constant factors. At test time, both \gls{drl} policies require a single actor forward pass plus the safety projection, yielding an inference complexity of $\mathcal{O}(P_a)$ per decision. In contrast, the AO-WMMSE baseline solves a sequence of matrix-valued subproblems with per-solve complexity on the order of $\mathcal{O}\big(a_{\max}(w_{\mathrm{in}} n_{\mathrm{bisect}} M^3 + N^3)\big)$, dominated by repeated $M \times M$ linear solves and an $N \times N$ eigendecomposition, while the robust AO-WMMSE-SAA variant preserves the same cubic scaling in $M$ and $N$ but adds a linear dependence on $S_{\mathrm{SAA}}$, the number of SAA scenarios, through the scenario aggregation, yielding $\mathcal{O}\big(a_{\max}(w_{\mathrm{in}} n_{\mathrm{bisect}} M^3 + S_{\mathrm{SAA}} N^3)\big)$.

Table~\ref{tab:complexity} reports the empirical offline and online runtimes for the considered methods under the parameter settings of Section~\ref{sec:hyperparameters}. In line with the above complexity discussion, the DRL agents incur a one-time offline training cost but achieve sub-millisecond inference, since deployment reduces to a single actor forward pass with the safety projection. In contrast, AO-WMMSE and AO-WMMSE-SAA require no training but take several hundred milliseconds per decision, which makes them impractical for fast time-varying \gls{uav} scenarios. In the proposed contextual bandit formulation, we omit target networks which lowers both memory usage and per-update overhead compared to standard \gls{ddpg} and \gls{td3} implementations and contributes to the higher training throughput observed in Table~\ref{tab:complexity}. Although TD3 maintains twin critics, its delayed actor updates partly offset the additional critic cost, and in our implementation it achieves slightly higher effective steps per second than \gls{ddpg}.

\begin{table}[htbp] \centering \caption{Computational Complexity Comparison} \label{tab:complexity} \begin{tabular}{@{}lrrr@{}} \toprule Algorithm & \begin{tabular}[c]{@{}r@{}}Offline Training \\ Time (s)\end{tabular} & \begin{tabular}[c]{@{}r@{}}Steps/s\end{tabular} & \begin{tabular}[c]{@{}r@{}}Online Inference \\ Time (ms)\end{tabular} \\ \midrule TD3 & 1227.2 & 162.97 & 0.62 \\ DDPG & 1336.6 & 149.63 & 0.62 \\ AO-WMMSE & 0.00 & 0.00 & 370.5 \\ AO-WMMSE-SAA & 0.00 & 0.00 & 551.01 \\ \bottomrule \end{tabular} \end{table}

\section{Conclusion}
This paper proposed a constrained, model-free \gls{drl} framework for joint \gls{bs} beamforming and \gls{ris} phase control in the presence of \gls{uav} attitude jitter and imperfect cascaded \gls{csi}. By formulating each channel realization as a contextual bandit and embedding a differentiable feasibility projection, the framework enables stable end-to-end learning while guaranteeing transmit power and unit-modulus constraints. We addressed uncertainty at training time by aligning the reward with the expected throughput objective. Also, we developed a \gls{ddpg} agent to achieve the objective, and a \gls{td3} agent to improve stability. Comprehensive experiments show that the learned agents perform competitively in both ideal and uncertain regimes, maintaining strong throughput relative to iterative \gls{ao}/\gls{wmmse} baselines while reducing online decision latency to the sub-millisecond scale. Future research directions include addressing stricter \gls{qos} constraints, incorporating a phase-dependent amplitude \gls{ris} model for more realistic implementation, extending to scenarios with multiple UAVs for increased coverage, and scaling to a larger set of \gls{ris} elements while maintaining proper hyperparameter tuning that preserves stability.

\section*{Acknowledgment}
We acknowledge the support provided by King Fahd University of Petroleum and Minerals (KFUPM), Dhahran 31261, Saudi Arabia, through the Interdisciplinary Research Center for Communication Systems and Sensing (IRC-CSS).
\bibliographystyle{IEEEtran}
\bibliography{Refs}

\end{document}